\documentclass[useAMS,usenatbib]{mn2e}
\usepackage[dvips]{graphicx}
\usepackage{amssymb}

\title[Escape velocity and mass in the outskirts of galaxy clusters]{Measuring the escape velocity and mass profiles of galaxy clusters beyond their virial radius}

\author[Serra et al.]{Ana Laura Serra $^{1,2}$\thanks{E-mail: serra@ph.unito.it}, Antonaldo Diaferio $^{1,2,3}$, 
Giuseppe Murante$^4$ 
\newauthor $\&$ Stefano Borgani$^{5,6,7}$ \\ 
$^1$Dipartimento di Fisica Generale ``Amedeo Avogadro'', Universit\`a degli Studi di Torino, Via P. Giuria 1, I-10125,
 Torino, Italy \\
$^2$ Istituto Nazionale di Fisica Nucleare (INFN), Sezione di Torino, Torino, Italy\\
$^3$ Harvard-Smithsonian Center for Astrophysics, MS20, 60 Garden St., Cambridge, MA 02138, USA\\ 
$^4$ INAF, Osservatorio Astronomico di Torino, Torino, Italy \\
$^5$ Dipartimento di Astronomia, Universit\`a di Trieste, Trieste, Italy \\
$^6$ INAF, Osservatorio Astronomico di Trieste, Trieste, Italy  \\
$^7$ Istituto Nazionale di Fisica Nucleare (INFN), Sezione di Trieste, Trieste, Italy \\
}

\begin{document}
\maketitle

\begin{abstract}
The caustic technique uses galaxy redshifts alone to measure the escape velocity and 
mass profiles of galaxy clusters to 
clustrocentric distances well beyond the virial radius, where dynamical equilibrium does not necessarily hold.
We provide a detailed description of this technique and analyse its possible systematic
errors. We apply the caustic technique to clusters with mass $M_{200}\ge 10^{14}h^{-1} M_\odot$ extracted
from a cosmological hydrodynamic
simulation of a $\Lambda$CDM universe. 
With a few tens of redshifts per squared comoving megaparsec within the cluster,
the caustic technique, on average, 
recovers the profile of the escape velocity from the cluster with better than $10$ percent accuracy up
to $r \sim 4 r_{200}$.
The caustic technique also recovers the mass profile with 
better than 10 percent accuracy in the range $(0.6-4)\, r_{200}$, but it  
overestimates the mass up to 70 percent at smaller radii.  
This overestimate is a consequence of neglecting the radial dependence of the 
filling function ${\cal F}_\beta(r)$.
The 1-$\sigma$ uncertainty on individual escape velocity profiles
increases from $\sim 20$ to $\sim 50$ percent when the radius 
increases from $r\sim 0.1 r_{200}$ to $\sim 4 r_{200}$.
Individual mass profiles have 1-$\sigma$ uncertainty between 40 and 80 percent
within the radial range $(0.6-4)\, r_{200}$. When the correct virial mass is known,
the  1-$\sigma$ uncertainty reduces to a constant 50 percent on the same radial range. 
We show that the amplitude of these uncertainties is completely due
to the assumption of spherical symmetry, which is difficult to drop. 
Other potential refinements of the technique are not crucial. 
We conclude that, when applied to individual clusters, the caustic technique
generally provides accurate escape velocity and mass profiles, although,
in some cases, the deviation from the real profile can be substantial. Alternatively, we can apply the technique to
synthetic clusters obtained by stacking individual clusters: in this case, 
the 1-$\sigma$ uncertainty on the escape velocity profile is smaller than 20 percent out to $4 r_{200}$.
The caustic technique thus provides reliable average profiles which extend to 
regions difficult or impossible to probe with other techniques. 
\end{abstract}


\section{Introduction}

Clusters of galaxies are valuable tools to measure 
the cosmological parameters and test structure formation models \citep[e.g.][]{voit05, diaf08}
and the galaxy-environment connection \citep[e.g.][]{ski09,hue09,mar08,mardom01}.
The evolution of the cluster abundance is a sensitive probe of the cosmological
parameters because clusters populate the exponential tail of the mass function of virialised
galaxy systems. Accurate mass measurements are however required
to avoid the propagation of systematic errors into the
estimation of the cosmological parameters.
There are two families of mass estimators: those which estimate the mass profiles
and those that measure the mass enclosed within a specific projected radius.

Traditionally, the estimation of the cluster mass is based on the
assumptions of spherical symmetry and dynamical equilibrium: either the cluster galaxies move
accordingly to the virial theorem \citep{zwicky37}, or the hot intracluster
plasma which emits in the X-ray band is in hydrostatic equilibrium 
within the gravitational potential well of the cluster \citep{sarazin88}.
More sophisticated approaches apply the Jeans equation for a steady-state
system, with the velocity anisotropy parameter $\beta$ as a further
unknown (\citealt{the86}; \citealt{merritt87};
see \citealt{biviano06} for a comprehensive
review of various improvements of this method).  

Dynamical equilibrium is also assumed when using 
the mass-X-ray temperature relation 
to estimate the cluster mass (e.g. \citealt{pierpaoli03}; see also
\citealt{borgani06}). Both the slope and the normalization of the
observed mass--temperature
relation is grossly reproduced by synthetic clusters obtained with
$N$-body/hydrodynamical simulations \citep{borgani04}. 
However, the presence of non--thermal pressure support (e.g., related
to turbulent gas motions) and the complex thermal structure of the
intra-cluster medium can significantly bias cluster mass estimates
based on the application of hydrostatic equilibrium to the X--ray
estimated temperature \citep[e.g.][]{rasia06,nagai07,piffaretti08}.
In principle, one can improve the mass estimate by exploiting the
integrated Sunyaev-Zel'dovich effect, which depends on the first power
of the gas density, rather than the square of the density as in the
X-ray emission, and yields a correlation with mass which is tighter
than the mass-X-ray temperature relation
\citep[e.g.][]{motl05,nagai06}. 
Only recently, such results have been confirmed by observations of 
real clusters \citep{Rines10, andersson10}.
However, their confirmation over a large statistical ensemble has to
await the results from ongoing large 
Sunyaev-Zel'dovich surveys. Alternatively, one can bypass
the problematic gas physics and use the correlation between mass and
optical richness, which is relatively easy to obtain from observations
in the optical/near-IR bands, but returns mass with a poorer accuracy 
\citep{andreon10}.

The dynamical equilibrium assumption can be dropped in the gravitational
lensing techniques, because the lensing effect depends only
on the amount of mass along the line of sight and not on its dynamical state
 \citep[e.g.][]{schneider}.
It is relevant to emphasize that all these methods do not measure the cluster
mass on the same scale: optical observations measure the mass within $\sim r_{200}$,
where $r_{200}$ is the radius within which the average mass density is
200 times the critical density of the universe; 
X-ray estimates rarely go beyond $\sim 0.5 r_{200}$, where the X-ray surface brightness
becomes smaller than the X-ray telescope sensitivity; lensing measures the central mass 
within $\sim 0.1 r_{200}$ or
the outer regions at radii larger than $\sim r_{200}$ depending on whether the strong or weak regime
applies. Scaling relations do not provide any information on the mass profile and 
give the total mass within a radius depending on the scaling
relation used, but still within $\sim r_{200}$.

\citet{diaferio97} (DG97, hereafter) 
suggested a novel method, the caustic technique, to estimate the mass 
from the central region out to well beyond $r_{200}$ with galaxy 
celestial coordinates and redshifts alone and 
without assuming dynamical equilibrium \citep{diaf09}. Prompted by the $N$-body simulations
of \citet{vanhaarlem93}, DG97 noticed that in hierarchical models of structure
formation, the velocity field in the cluster outskirts is not
perfectly radial, as expected in the spherical infall model (\citealt{regos89}; \citealt{hiotelis01}) 
but has a substantial random component. They thus suggested
to exploit this fact 
to extract the galaxy escape velocities as a function
of radius from the distribution of galaxies in redshift space.
In the caustic method, the velocity anisotropy parameter
$\beta$ and the mass density profile in hierarchical clustering models
combine in such a way that their knowledge is largely unnecessary in estimating
the mass profile. This property explains the power of the method.  

This method is particularly relevant because it is an alternative to
lensing to measure mass in the cluster external regions and, unlike
lensing, it can be applied to clusters at any redshift, provided there
are enough galaxies to sample the redshift diagram properly. We will
see below that a few tens of redshifts per squared comoving megaparsec
within the cluster are sufficient to apply the technique
reliably. This request might have appeared demanding a decade ago, but
it is perfectly feasible for the large redshift surveys currently
available. 

\citet*{geller99} were the first to apply the caustic method: 
they measured the mass profile of Coma out to $10 h^{-1}$ Mpc
from the cluster centre and were able to demonstrate that the
\citet*{nfw} (NFW) profile fits well the cluster density profile
out to these very large radii, thus ruling out the isothermal sphere
as a viable cluster model; a few years later, the failure of the
isothermal model was confirmed by the first analysis based on
gravitational lensing applied to Cl~0024 \citep{kneib03}.  The
goodness of the NFW fit out to $5-10 h^{-1}$ Mpc was confirmed by
applying the caustic technique to a sample of nine clusters densely
sampled in their outer regions, the Cluster And Infall Region Nearby
Survey (CAIRNS, \citealt{rines03}), and later to a complete sample of
72 X-ray selected clusters with galaxy redshifts extracted from the
Fourth Data Release of the Sloan Digital Sky Survey (Cluster Infall
Regions in the Sloan Digital Sky Survey: CIRS, \citealt{rines06a}) and
from the Fifth Data Release \citep{rines10b}.  CIRS is currently the
largest sample of clusters whose mass profiles have been measured out
to $\sim 3 r_{200}$; \citet{rines06a} were thus able to obtain a
statistically significant estimate of the ratio between the masses
within the infall and the virial regions: they found a value of
$2.2\pm 0.2$, in agreement with current models of cluster
formation. These analyses have been extended to a sample of groups of
galaxies \citep{rines10b}. \citet*{rines06b} also used the CIRS sample
to estimate the virial mass function of nearby clusters and determined
cosmological parameters consistent with WMAP values \citep{dun09}.

A good fit with the NFW profile out to $\sim 2 r_{200}$ was also found
by \citet{biviano03} who applied the caustic technique to an ensemble
cluster obtained by stacking 43 clusters from the Two Degree Galaxy
Redshift Survey (2dGFRS, \citealt{colless01}).  Here, unlike the
previous analyses, the caustic method was not applied to individual
clusters because the number of galaxies per cluster was relatively
small.  Recently, \cite{lem09} have applied both the caustic technique
and the Jeans analysis to $\sim 500$ cluster members of A1689.  The
estimated virial mass from both methods agrees with previous lensing
and X-ray analyses.

The caustic method does not rely on the dynamical state of the cluster
and of its external regions: there are therefore estimates of the mass
of unrelaxed systems, namely the Shapley supercluster
\citep{reisenegger00,pro06}, the Fornax poor cluster, which shows two
distinct dynamical components \citep*{drinkwater01}, the A2199 complex
\citep{rines02}, and two clusters, A168 and A1367, which are
undergoing major mergers \citep{rines03}.
More recently, \citet{diaf05b} analysed Cl~0024, a cluster that is
likely to have suffered a recent merging event \citep{czoske02}, and
showed that the caustic mass and the lensing mass agree with each
other, but disagree with the X-ray mass, which is the only estimate
relying on dynamical equilibrium.

The method was shown to return reliable mass profiles 
when applied to synthetic clusters extracted from $N$-body simulations.
However, these analyses of the performance of the method either were done when
the operative procedure of the technique was not yet completely
developed (DG97) or the results of a fully detailed analysis of the systematic errors
was not provided \citep{diaf99} (D99, hereafter). The dense redshift surveys currently available,
especially around clusters, provide ideal data sets where the caustic
technique can now be applied.  With this perspective, it is therefore
timely to reconsider the possible systematic errors and biases of this
technique, in view of a robust application to the new data sets.

The purpose of this paper is to provide both a transparent description
of the technique, with some improvements to what is described in
D99, and a thorough statistical analysis.
The caustic technique represents a powerful method to infer
cluster mass profiles and complements other 
methods based on X-ray, Sunyaev-Zeldovich and lensing observations
which probe different scales of the clusters.

In Section \ref{sec:basics} we describe the basic idea of the caustic
technique. Section \ref{sec:sample} describes the simulated cluster
sample used, and Section \ref{sec:CT} describes the implementation of
the technique.  In Sections \ref{sec:GP} and \ref{sec:mass} we
estimate the accuracy of the escape velocity and mass profiles of
galaxy clusters estimated with the caustic technique.  In Section
\ref{sec:system} we investigate the systematics due to the choice of
the parameters.
We summarize our main results and draw our conclusions in Section
\ref{sec:concl}. 

\section{Basics}\label{sec:basics}

In this section, we briefly review the simple physical idea behind the
caustic technique. More details are given in DG97 and D99.

In hierarchical clustering models of structure formation, clusters
form by the aggregation of smaller systems falling onto the cluster
from the surrounding region. The accretion does not take place purely
radially \citep[e.g.][]{white2010}, therefore galaxies or dark matter particles within the
falling clumps have velocities with a substantial non-radial
component. Specifically, the r.m.s. $\langle v^2\rangle$ of these
velocities is due to the gravitational potential of the cluster and
the groups where the galaxy resides, and to the tidal fields of the
surrounding region. When viewed in the redshift diagram, viz. the
plane of the line-of-sight (l.o.s.) velocity w.r.t. the cluster centre
and the clustrocentric radius $r$, galaxies populate a region with a
characteristic trumpet shape with decreasing amplitude ${\cal A}$ with
increasing $r$.  This amplitude is related to $\langle
v^2\rangle$. The breakthrough of DG97 was to identify this amplitude
with the escape velocity from the cluster region corrected for a
function depending on the velocity anisotropy parameter $\beta$.

Assuming a spherically symmetric system, the escape velocity $v_{\rm esc}^2(r)=-2\phi(r)$, where $\phi(r)$ is the
gravitational potential originated by the cluster, is a non-increasing function of $r$, 
because gravity is always attractive and ${\rm d}\phi/{\rm d}r>0$. At any
given radius $r$, we expect that observing a galaxy with
a velocity larger than the escape velocity is unlikely. Thus, we identify the escape velocity with
the maximum velocity that can be observed. It follows that the amplitude
${\cal A}$ at the projected radius $r_\perp$ measures the average
component along the l.o.s. of the escape velocity at the three-dimensional radius $r=r_\perp$.
To determine this average component of the velocity, we consider the velocity 
anisotropy parameter 
$\beta(r)=1-(\langle v^2_\theta\rangle + \langle v^2_\phi\rangle)/2\langle v^2_r\rangle $,
where $v_\theta$, $v_\phi$, and $v_r$ are the longitudinal, azimuthal and 
radial components of the velocity ${\mathbf v}$ of a galaxy, respectively, and the brackets
indicate an average over the velocities of the galaxies in the volume $\mbox{d}^3{\bf r}$
centred on position ${\bf r}$.
If the cluster rotation is negligible, $\langle v^2_\theta\rangle=\langle v^2_\phi\rangle=\langle v^2_{\rm los}\rangle$,
where $v_{\rm los}$ is the l.o.s. component of the velocity, and 
$\langle v^2_r\rangle=\langle v^2\rangle-2\langle v^2_{\rm los}\rangle$. By substituting
in the definition of $\beta$, we obtain $\langle v^2\rangle=\langle v^2_{\rm los} \rangle g(\beta)$ where
\begin{equation}
g(\beta) = {3-2\beta(r)\over 1-\beta(r)}\; .
\end{equation}
By applying this relation
to the escape velocity at radius $r$, $\langle v_{\rm esc}^2(r)\rangle=-2\phi(r)$,
and assuming that ${\cal A}^2(r)=\langle v^2_{\rm esc, los}\rangle$, 
we obtain the fundamental relation between the gravitational potential $\phi(r)$ and
the observable caustic amplitude ${\cal A}(r)$
\begin{equation}
-2\phi(r)={\cal A}^2(r)g(\beta) \equiv \phi_\beta(r) g(\beta) \; .
\label{eq:rig-pot}
\end{equation}
Note that the gravitational potential profile is related to the
caustic amplitude by the function $g(\beta)$. Therefore, after the
caustic amplitude estimation, $\beta$ becomes the only unknown
function for the gravitational potential estimation.

The further novel suggestion of DG97 is to use this relation to infer the cluster mass to very large radii.
To do so, one first notices that the mass of an infinitesimal shell can be
cast in the form $G \,\mbox{d}m=-2\phi(r){\cal F}(r) \,\mbox{d}r = {\cal A}^2(r)g(\beta) {\cal F}(r) \,\mbox{d}r$ where 
\begin{equation}
{\cal F}(r)=-2\pi G{\rho(r)r^2\over \phi(r)}\; .
\end{equation}
Therefore the mass profile is
\begin{equation}
GM(<r)=\int_0^r {\cal A}^2(r) {\cal F}_\beta(r) \,\mbox{d}r
\label{eq:rig-massprof}
\end{equation}
where ${\cal F}_\beta(r) = {\cal F}(r) g(\beta)$.

Equation (\ref{eq:rig-massprof}) however only relates the 
mass profile to the density profile of a spherical system and a profile
can not be inferred without knowing the other. DG97 solve this 
impasse by noticing that in hierarchical clustering
scenarios, ${\cal F}(r)$ is not a strong function of $r$. This is easily seen 
in the case of the NFW model which is an excellent description of dark matter
density profiles in the hierarchical clustering scenario:
\begin{equation}
{\cal F}_{\rm NFW}(r) = {r^2\over 2(r+r_s)^2}{1\over \ln(1+r/r_s)} \; ,
\label{eq:Fnfw}
\end{equation}
where $r_s$ is a scale factor.
One expects that ${\cal F}_\beta(r)$ is also a slowly changing function of $r$ if
$g(\beta)$ is. DG97 and D99 show that this is the case, and here
we confirm their result. The final, somewhat strong, assumption is then to consider ${\cal F}_\beta(r)=
{\cal F}_\beta={\rm const}$
altogether and adopt the recipe
\begin{equation}
GM(<r)={\cal F}_\beta\int_0^r {\cal A}^2(r) \,\mbox{d}r \; .
\label{eq:recipe-massprof}
\end{equation}

It is appropriate to emphasize that equations (\ref{eq:rig-pot}) and (\ref{eq:rig-massprof}) are 
rigorously correct, whereas
equation (\ref{eq:recipe-massprof}) is a heuristic recipe to estimate the mass profile. Based
on equation (\ref{eq:rig-pot}) the caustic technique can also estimate a combination of 
the  gravitational potential profile and velocity anisotropy parameter $\beta$. Below, we show the accuracy of
these estimates.

\section{The simulated cluster sample}\label{sec:sample}

Before analysing the systematic errors of the caustic technique, we need to
check how well the physical assumptions of the method are satisfied.
We do so by using $N$-body simulations, assuming that these
simulations are a faithful representation of the real universe.

Here we use the cluster sample extracted from the cosmological
$N$-body/hydrodynamical simulation described in \citet{borgani04}.
They simulate a cubic volume, 192 $h^{-1}$ Mpc on a side, of a flat
$\Lambda$CDM universe, with matter density $\Omega_0=0.3$, Hubble
parameter $H_0=100 h$ km s$^{-1}$ Mpc$^{-1}$ with $h=0.7$, power
spectrum normalization $\sigma_8=0.8$, and baryon density $\Omega_{\rm
  b}=0.02 h^{-2}$.  The density field is sampled with $480^3$ dark
matter particles and an initially equal number of gas particles, with
masses $m_{\rm DM}=4.6\times 10^9 h^{-1} M_\odot$ and $m_{\rm gas}=
6.9\times 10^8 h^{-1} M_\odot$, respectively. The Plummer-equivalent
gravitational softening is 
set to 7.5 $h^{-1}$ physical kpc at $z<2$, while being fixed in
comoving units at higher redshift.

The simulation was run with {\tt GADGET-2} \citep*{springel01},
a massively parallel Tree+SPH code with fully adaptive time--stepping.
Besides standard hydrodynamics, it also contains a treatment of
radiative gas cooling, star formation in a multi-phase inter-stellar
medium and feedback from supernovae in the form of galactic ejecta 
\citep[e.g.,][]{borgani04,diaf05a}. 
Here we are only interested in the gravitational dynamics of the
matter distribution.

The simulation volume yields a cluster sample large enough for
statistical purposes.  We identify clusters in the simulation box with
a two-step procedure: a friends-of-friends algorithm applied to the
dark matter particles alone provides a list of halos whose centres are
used as input to the spherical overdensity algorithm which outputs the
final list of clusters \citep{borgani04}. Centred on the most bound
particle of each cluster, the sphere with virial overdensity 200, with
respect to the critical density, defines the virial radius
$r_{200}$. At redshift $z=0$ the simulation box contains 
100 clusters with mass $M(<r_{200})\ge 10^{14} h^{-1} M_\odot$.  This
cluster set is our reference sample.

\subsection{Cluster properties in 3D}

\begin{figure}
\includegraphics[angle=0,scale=.58, bb=70 25 584 310]{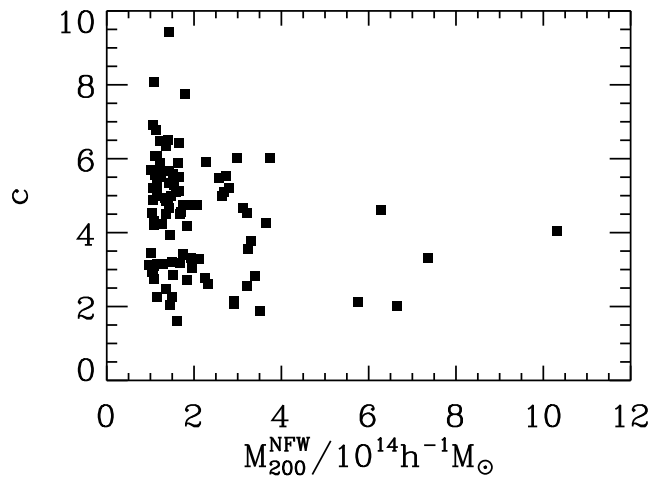}
\caption{Concentration parameters $c$ vs. the cluster mass $M_{200}^{\rm NFW}$ of our simulated
cluster sample.}
\label{fig:c-vs-mass}
\end{figure}

\begin{figure}
\includegraphics[angle=0,scale=.58,bb=60 20 554 310]{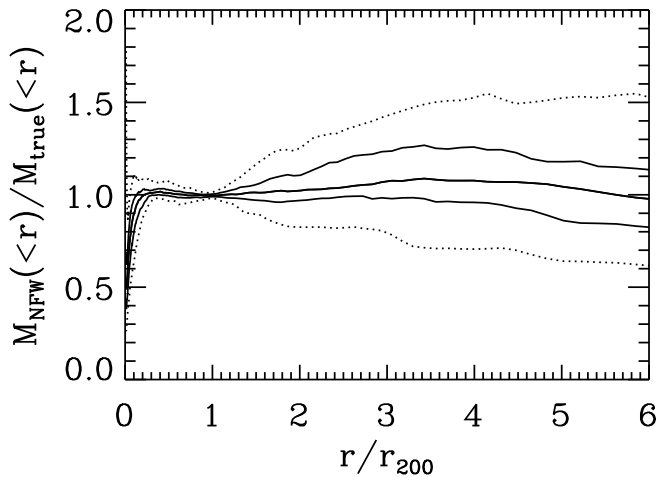}
\caption{Profiles of the ratio between the mass profile of each
  cluster predicted by the NFW fit and its true mass profile: 90 (50)
  percent of the profiles are within the upper and lower dotted
  (solid) curves.  The central solid curve is the median profile. For
  each cluster, the NFW fit is only performed to the mass distribution
  within $1 h^{-1}$ Mpc.}
\label{fig:mprof3D}
\end{figure}

\begin{figure}
\includegraphics[angle=0,scale=.58,bb=60 20 554 327]{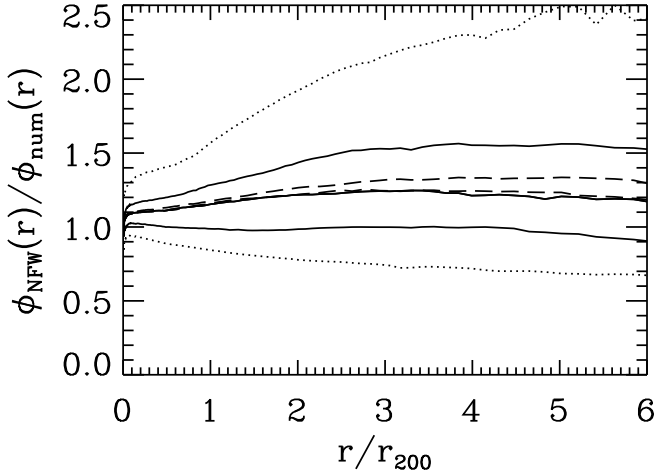}
\caption{Profiles of the ratio between the gravitational potential
  profile of each cluster predicted by the NFW fit and the numerical
  gravitational potential profile derived from the true mass
  distribution within $r_{\rm max}=10 h^{-1}$ Mpc from the cluster
  centre: 90 (50) percent of the profiles are within the upper and
  lower dotted (solid) curves.  The central solid curve is the median
  profile. The lower and upper dashed curves are the median profiles
  for $r_{\rm max}=8$ and $6 h^{-1}$ Mpc, respectively.}
\label{fig:pot3D}
\end{figure}

We fit the three-dimensional (3D) real cumulative mass profile with
the NFW model
\begin{equation}
M(<r) = M_s \left[ \ln\left(1+{r\over r_s}\right) - {r/r_s\over 1+r/r_s}\right]
\end{equation} 
where $M_s=M(<r_s)/(\ln 2-1/2)=4\pi \delta_c \rho_{\rm cr} r_s^3$, $\delta_c=(200/3)c^3/[\ln(1+c)-c/(1+c)]$, and
$c=r_{200}^{\rm NFW}/r_s$. We fit the mass profile rather than the density profile to resemble
the procedure adopted with real clusters \citep[e.g.][]{diaf05b}. Similarly, for the fit, we only consider
the mass distribution within $r=1 h^{-1}$ Mpc. Moreover, 
the parameters $M_s$ and $r_s$ are less correlated than $\delta_c$ and $r_s$ and
adopting them in the fit provides more robust results \citep{mahdavi99}.

From the NFW fit we derive the parameters $r_{200}^{\rm NFW}$ and
$c$. The relation between $c$ and $M_{200}^{\rm NFW}=(4\pi/3)
(r_{200}^{\rm NFW})^3 200\rho_{\rm cr}$ is shown in Figure
\ref{fig:c-vs-mass}.  Our cluster sample provides the percentile
ranges $r_{200}^{\rm NFW}=[0.77,0.86,1.17]h^{-1}{\rm Mpc} $ and
$c=[2.06,4.62,6.50]$.\footnote{Throughout this paper the notation
  $[q_1,q_2,q_3]$ shows the median $q_2$ and the range $[q_1,q_3]$
  which contains 90 percent of the sample.}  The radius $r_{200}^{\rm
  NFW}$ is basically identical to the true $r_{200}$ derived from the
actual mass distribution. Their ratio is $r_{200}^{\rm
  NFW}/r_{200}=[0.99,1.00,1.02]$.  The NFW model is an excellent fit
to the true mass profile (see Figure \ref{fig:mprof3D}) within
$r_{200}$. Therefore the ratio between the mass $M_{200}^{\rm NFW}$
and the actual $M_{200}$ is correspondingly close to 1: $M_{200}^{\rm
  NFW}/M_{200}=[0.96,1.00,1.05]$.
The NFW fit is on average very good also at radii larger than $r_{200}$, although
the scatter substantially increases. In the very centre, the NFW model underestimates
the mass by 50 percent. This is due to the fact that the total mass includes dark matter, gas,
and stars and in the simulation there is always a central galaxy which is
unrealistically massive: the stellar particles, on average,  
contribute 50 (15) percent of the total mass within 0.02 (0.1) $r_{200}$. When the NFW fitting routine
tries to accomodate the mass profile within $1 h^{-1}$ Mpc, underestimating the
mass profile in this very central region yields the minimum $\chi^2$.  

The caustic amplitude is related to the gravitational potential $\phi(r)$ and we are thus 
interested in determining $\phi(r)$ in our simulation.
For an isolated spherical system with density profile $\rho(r)$, the potential $\phi(r)$
obeying the Poisson equation can be cast in the form  
\begin{equation}
\phi(r) = -4\pi G\left[{1\over r}\int_0^r \rho(x) x^2 \,\mbox{d}x + \int_r^\infty \rho(x) x \,\mbox{d}x\right]\; .
\label{eq:phi}
\end{equation}
In the real universe, the relevant quantity is the gravitational
potential originating from the mass density fluctuations around the
mean density $\langle \rho \rangle$. We can thus use equation
(\ref{eq:phi}) for non-isolated clusters by replacing $\rho(r)$ with $
\rho(r)-\langle\rho\rangle \equiv \langle\rho\rangle\delta(r)$.  The
second integral is finite, because at sufficiently large $r$,
$\delta(r)\sim 0$.  In the simulation, we replace the upper limit of the
second integral with $r_{\rm max}=10 h^{-1}$ Mpc. Figure \ref{fig:pot3D}
compares $\phi_{\rm num}(r)$ computed from the actual mass
distribution around each cluster in the simulation, with the
gravitational potential
\begin{equation}
\phi_{\rm NFW}(r) = -{GM_s\over r}  \ln \left(1+{r\over r_s}\right)
\end{equation}
expected from an isolated cluster described by the NFW model.  The NFW
potential well is deeper than the numerical $\phi_{\rm num}(r)$,
because it neglects the mass surrounding the cluster. This mass 
exerts a pull to the mass within the
cluster that makes the actual $\phi_{\rm num}(r)$ to be 10--30 percent
shallower.  The upper limit $r_{\rm max}$ of the second integral of
equation (\ref{eq:phi}), adopted in the numerical estimate, plays a
negligible effect when it is large enough: the median $\phi_{\rm
  num}(r)$ for $r_{\rm max}=8 h^{-1}$ Mpc is indistinguishable from
the profile computed with $r_{\rm max}=10 h^{-1}$ Mpc.

\begin{figure*}
\includegraphics[angle=0,scale=.8]{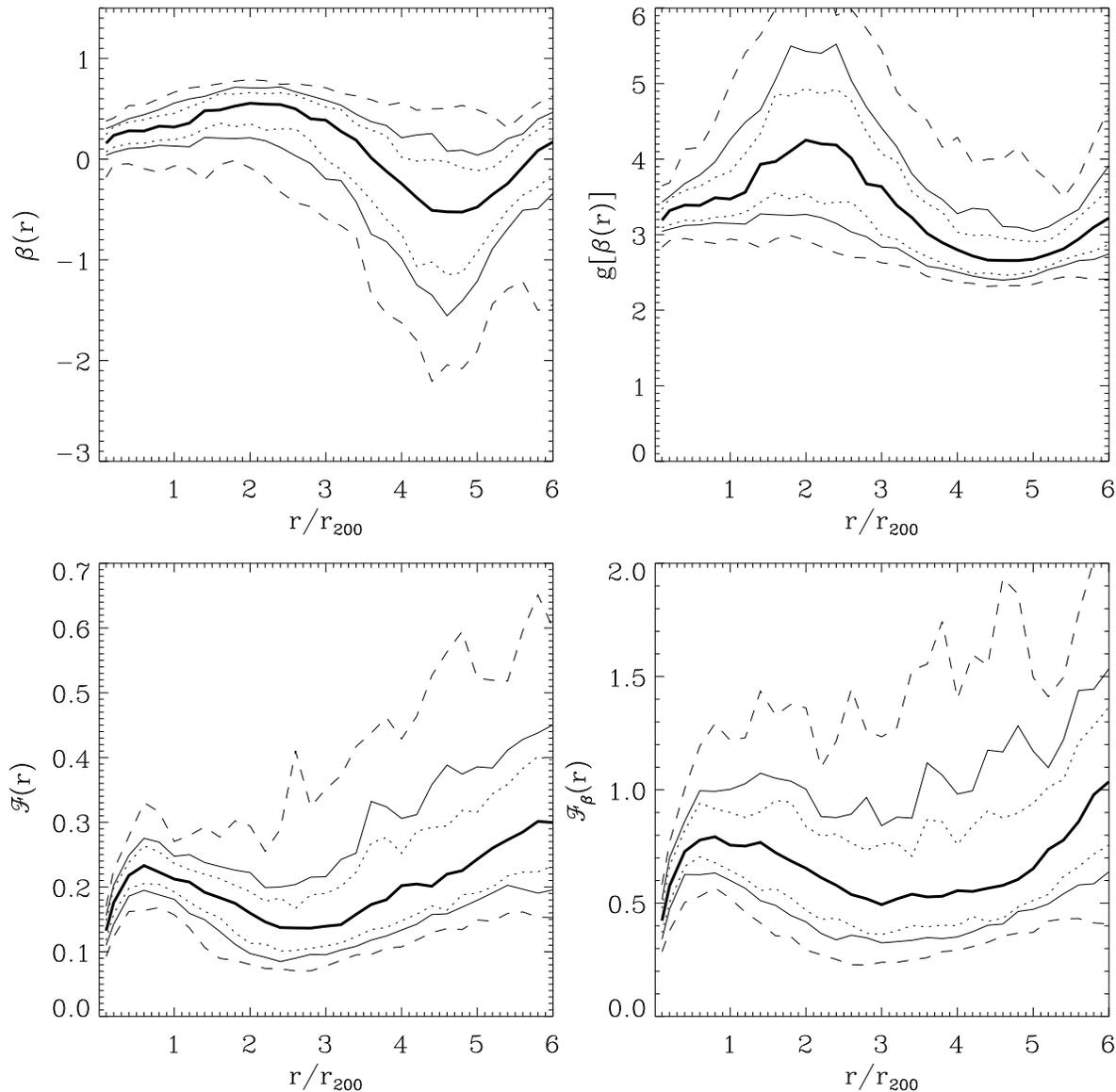}
\caption{Profiles of the functions $\beta(r)$, $g(\beta)$, ${\cal F}(r)$ and ${\cal F}_\beta(r)$ described
in the text: 50, 68 and 90 percent of the profiles are within the upper and lower dotted, solid and dashed curves.
The central solid curves are the median profiles.}   
\label{fig:beta-4pan}
\end{figure*}

Figure \ref{fig:beta-4pan} shows the profiles of the velocity
anisotropy parameter $\beta(r)$ and the other functions defined in
Section \ref{sec:basics} for our sample of 100 simulated galaxy
clusters.  This figure confirms the results of the $\Lambda$CDM model
presented in Figure 3 of D99 (see also Figure 25 in \citealt{diaf01},
which shows the velocity field of simulated galaxies rather than dark
matter particles).  Specifically, Figure \ref{fig:beta-4pan} shows
that, on average, $\beta(r)\la 0.7$ at $r\la 3r_{200}$ and
consequently, on these scales, the median profile of $g(\beta)$ varies
by less than 30 percent. Individual clusters may of course have wider
variations.

We remind that $\beta(r)$ is derived from the velocities of all the
particles in the simulations (dark matter, gas and stars). These
velocities are set in the rest frame of each cluster and are augmented
of the Hubble flow contribution $H_0{\bf r}$.  This term provides an
increasing contribution to the particle velocity which is not
negligible at $r\ga (1-2) r_{200}$. Specifically, by assuming
dynamical equilibrium and setting $v_p^2(r)\sim GM(<r)/r$, where
$v_p(r)$ is the average peculiar velocity at $r$, we see that $v_p$
decreases with $r$; therefore, the total velocity ${\bf v}(r)={\bf
  v}_p(r)+H_0{\bf r}$ reaches a minimum value when $v_p(r)\sim H_0 r$,
viz. when $r\sim [GM(<r)/H_0^2]^{1/3}\sim 4.01 h^{-1}$ Mpc, where we
have set $M=1.49\times 10^{14} h^{-1} M_\odot$, the median $M_{200}$
of our cluster sample. Since the tangential component of the total
velocity is unaffected by the Hubble flow, the minimum of $v_p(r)$,
and consequently of $\langle v_r^2\rangle$, is a minimum of
$\beta$. For our sample the median $r_{200}=0.86 h^{-1}$ Mpc, and we
have a minimum $\beta$ at $r/r_{200}=4.66$ which is in rough agreement
with Figure \ref{fig:beta-4pan}.

Finally, Figure \ref{fig:beta-4pan} shows that ${\cal F}(r)$ and ${\cal F}_\beta(r)$ are slowly varying functions of $r$, 
as expected. 
We discuss these profiles in Section \ref{sec:mass}.

\subsection{The mock cluster catalogue}\label{subsec:mockcatalogue}

We compile mock redshift catalogues from our 100 simulated clusters as
follows.  We project each cluster along ten random lines of sight; for
each line of sight, we also project the clusters along two additional
lines of sight in order to have a set of three orthogonal projections
for each random direction. We thus have a total of $100\times 10
\times 3=3000$ mock redshift surveys. We locate the cluster centre at
the celestial coordinate $(\alpha,\delta)=(6^h,0^\circ)$ and redshift
$cz=32000$ km~s$^{-1}$. We consider a field of view of $12 h^{-1}$~Mpc 
on a side at the cluster distance.  The simulation box is $192 
h^{-1}$~Mpc on a side, and the mock survey contains a large fraction
of foreground and background large-scale structure.  We only consider
a random sample of 1000 dark matter particles in each mock catalogue.
Only a fraction of these 1000 particles are within $3r_{200}$ of the
cluster centre, specifically this number has percentile range
$[96,185,408]$ in our 3000 mock catalogues.  In
Sect. \ref{subsec:noofparticles} we investigate the effect of varying
the number of particles in the mock survey.

\section{The caustic Technique}\label{sec:CT}

The celestial coordinates and redshifts of the galaxies within the
cluster field of view are the input data of the caustic technique. The
technique proceeds along four major steps:
\begin{enumerate}
\renewcommand{\theenumi}{(\arabic{enumi})}
\item arrange the galaxies in a binary tree according to a hierarchical method;
\item select two thresholds to cut the tree twice, at an upper and a
  lower level: the largest group obtained from the upper-level
  threshold identifies the cluster candidate members; the lower-level
  threshold identifies the substructures of the cluster; the cluster
  candidate members determine the cluster centre, its velocity
  dispersion and size;
\item with the cluster centre of the candidate members, build the
  redshift diagram of all the galaxies in the field; with the velocity
  dispersion and size of the candidate members determine the threshold
  $\kappa$ which enters the caustic equation, locate the caustics, and
  identify the final cluster members;
\item the caustic amplitude determines the escape velocity and mass profiles.
\end{enumerate}
We detail these steps in turn.

\subsection{Binary tree}

\begin{figure*}
\includegraphics[angle=0,scale=.85,bb=70 100 680 550]{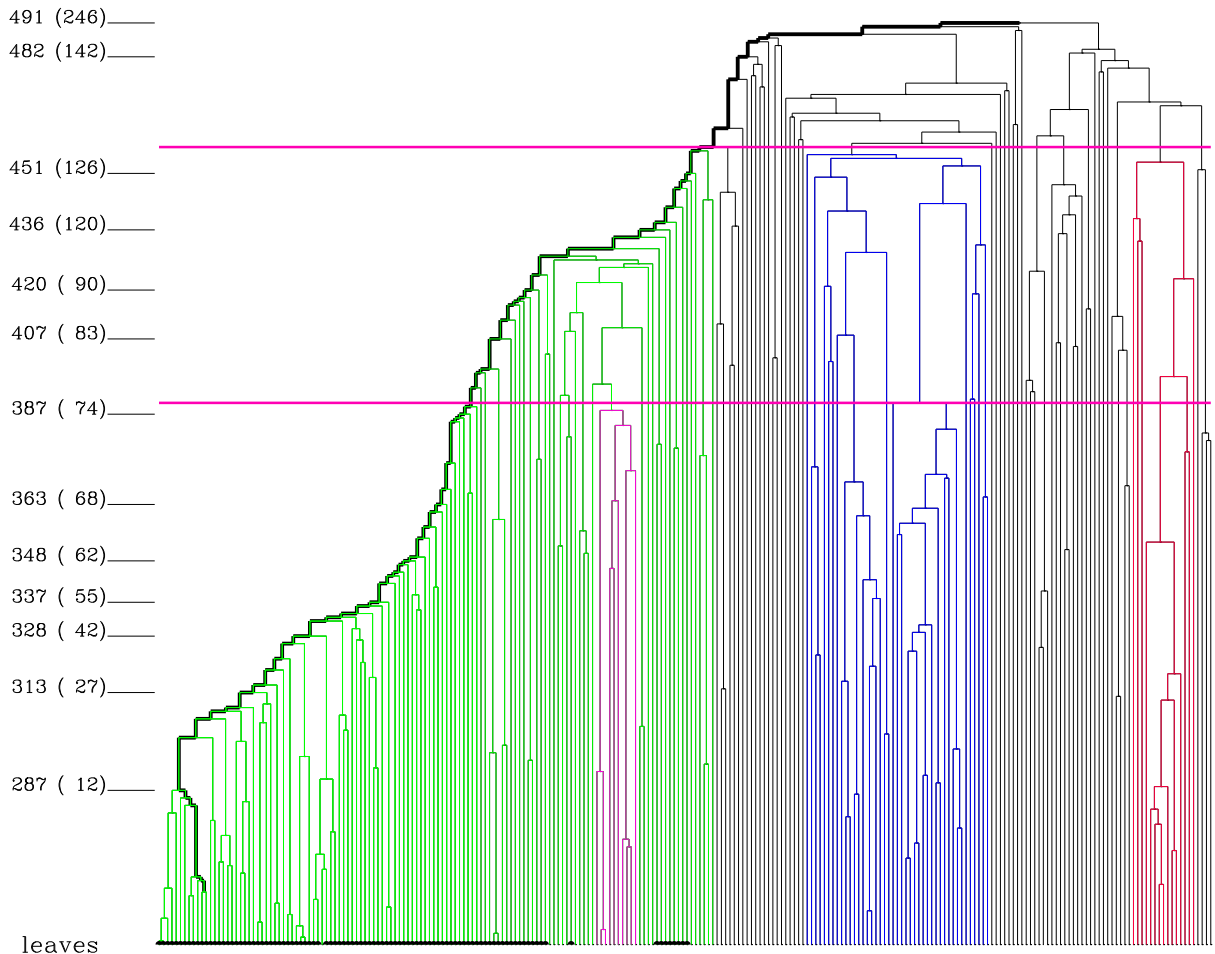}
\caption{Dendrogram representation of the binary tree of a  subsample of the particles in the field of a simulated cluster. 
The particles are the leaves of the tree at the bottom of the plot. The particles within $3r_{200}$ in real space are highlighted in black. The thick path
highlights the main branch. The horizontal lines show the upper
and lower thresholds used to cut the tree. Only as a guide, some nodes are labelled on the left-hand side, with their 
number of descendants in brackets.} 
\label{fig:dendrogram}
\end{figure*}

Each galaxy is located by the vector ${\bmath
  r}_i=(\alpha_i,\delta_i,r_i)$, where $\alpha_i$ and $\delta_i$ are
the celestial coordinates, and $r_i$ is the comoving distance to the observer 
of a source at redshift $z_i$, i.e. the spatial part of the geodesic
travelled by a light signal from the source to the observer. The
comoving radial coordinate $r_i$ is determined by the relation
$\int_0^{r_i} dx/\sqrt{1-\kappa x^2} = \int_0^{z_i} cdz/H(z)$, where
$\kappa = (H_0/c)^2(\Omega_0+ \Omega_\Lambda -1) $ is the space
curvature and $H^2(z) = H_0^2
[\Omega_0(1+z)^3+(1-\Omega_0-\Omega_\Lambda)(1+z)^2 + \Omega_\Lambda]$
is the Hubble constant at redshift $z$.  In the Einstein-de Sitter
universe ($\kappa=0$, $\Omega_0=1$)
\begin{equation}
r_i = {2c\over H_0}\left[1-{1\over (1+z_i)^{1/2}}\right]\; .
\end{equation}
The comoving separation $r_{12}$ between two sources with comoving
radial coordinates $r_1$ and $r_2$ and angular separation 
$\theta$ as observed from $O$ is 
\begin{eqnarray}
r_{12}^2 & =& r_1^2+r_2^2 - \kappa r_1^2r_2^2(1+\cos^2\theta)- 2r_1r_2\cos\theta\times\cr
 & & \times \sqrt{1-\kappa r_1^2}\sqrt{1-\kappa r_2^2}\; .
\label{eq:r12}
\end{eqnarray}
This equation is a generalization of the cosine law on the two-dimensional 
surface of a sphere (e. g. \citealt{peebles93}, equation 
12.42).

For the construction of the binary tree, we are interested in defining
an appropriate similarity based on the pairwise separation $r_{12}$.
In general, the rank of the separation $r_{12}$ of a number of
sources with different redshift $z_i$ is not preserved when we vary
$\Omega_0$ and $\Omega_\Lambda$. However, a sufficient condition for
having the ranks preserved is that the universe is never collapsing
and $\kappa\le 0$. This condition is satisfied when $\Omega_0\le 1$
and $\Omega_0+\Omega_\Lambda\le 1$. 

Based on this argument, we can compute $r_{12}$ in the Einstein-de Sitter  universe, and use the
ordinary Euclidean geometry to derive the binary
tree similarity. For each galaxy pair,  
we define the l.o.s. vector ${\bmath l}=({\bmath r}_1+{\bmath r}_2)/2$
and the separation vector ${\bmath r}_{12}={\bmath r}_1-{\bmath r}_2$. 
We then determine the component of ${\bmath r}_{12}$ along the
line of sight, 
\begin{equation}
\pi={{\bmath r}_{12}\cdot {\bmath l}\over \vert {\bmath l}\vert}=
{r_1^2-r_2^2\over\vert{\bmath r}_1+{\bmath r}_2\vert}\; ,
\label{eq:pi}
\end{equation} 
and the component perpendicular to the line of sight 
\begin{equation}
r_p=(r_{12}^2-\pi^2)^{1/2}.
\label{eq:rp}
\end{equation} 
We now consider the proper l.o.s. velocity difference $\Pi=\pi/(1+z_l)$ and the
proper projected spatial separation $R_p=r_p/(1+z_l)$ of the
galaxy pair at the intermediate redshift $z_l$ that satisfies the
relation $r_l=(r_1+r_2)/2$. We adopt the similarity 
\begin{equation}
E_{ij}=-G{m_i m_j\over R_p}+{1\over 2}{m_i m_j\over m_i+m_j}\Pi^2 \; .
\label{eq:pairwise-energy}
\end{equation}
This similarity reduces to the one adopted by D99 at small $z$.
In equation (\ref{eq:pairwise-energy}), the galaxy mass is a free parameter.
D99 assumes $m_i=m_j=10^{12}h^{-1}$ M$_\odot$,
as we do here. Varying the galaxy mass changes the details 
of the binary tree, because one differently weights $\Pi$ and $R_p$. However, in the range $(10^{11},10^{13})h^{-1}$ M$_\odot$,
the centre location and the galaxy membership remain substantially unchanged. One can of course include different galaxy masses proportional to the galaxy luminosities.
This prescription turns out to be unnecessary, because it would not increase the accuracy of 
the determination of the mass and escape velocity profiles.
In fact, we will see below (Section \ref{sec:projeff}) that projection effects completely dominate
the uncertainties of the caustic method.

For the sake of completeness, we note that, in a non-flat universe,
with $\Omega_0< 1$ and $\Omega_0+\Omega_\Lambda< 1$,
one can still perform the component separation
with equations (\ref{eq:pi}) and (\ref{eq:rp}) on a flat space tangent 
to the observer's location point $O$, provided one defines an appropriate 
bi-unique function between a geodesic distance $r_i$ and a
vector ${\bmath r}_i$ on the flat space. 

The binary tree is built as follows: (i) each galaxy can be thought as a group $g_\alpha$; (ii) to each
group pair $g_\alpha, g_\beta$ it is associated the binding energy $E_{\alpha\beta}={\rm min}\{E_{ij}\}$, 
where $E_{ij}$ is the binding energy between the galaxy $i\in g_\alpha$ and the galaxy $j\in g_\beta$;
(iii) the two groups with the smallest binding energy $E_{\alpha\beta}$ are replaced with
a single group $g_\gamma$ and the total number of groups is decreased by one; (iv) the
procedure is repeated from step (ii) until we are left with only one group.
Figure \ref{fig:dendrogram} shows the binary tree of a simulated cluster as an example.

\subsection{Cutting the binary tree: The $\sigma$ plateau}\label{subsec:sigmaplateau}

\begin{figure}
\includegraphics[angle=0,scale=.47,bb=30 14 566 390]{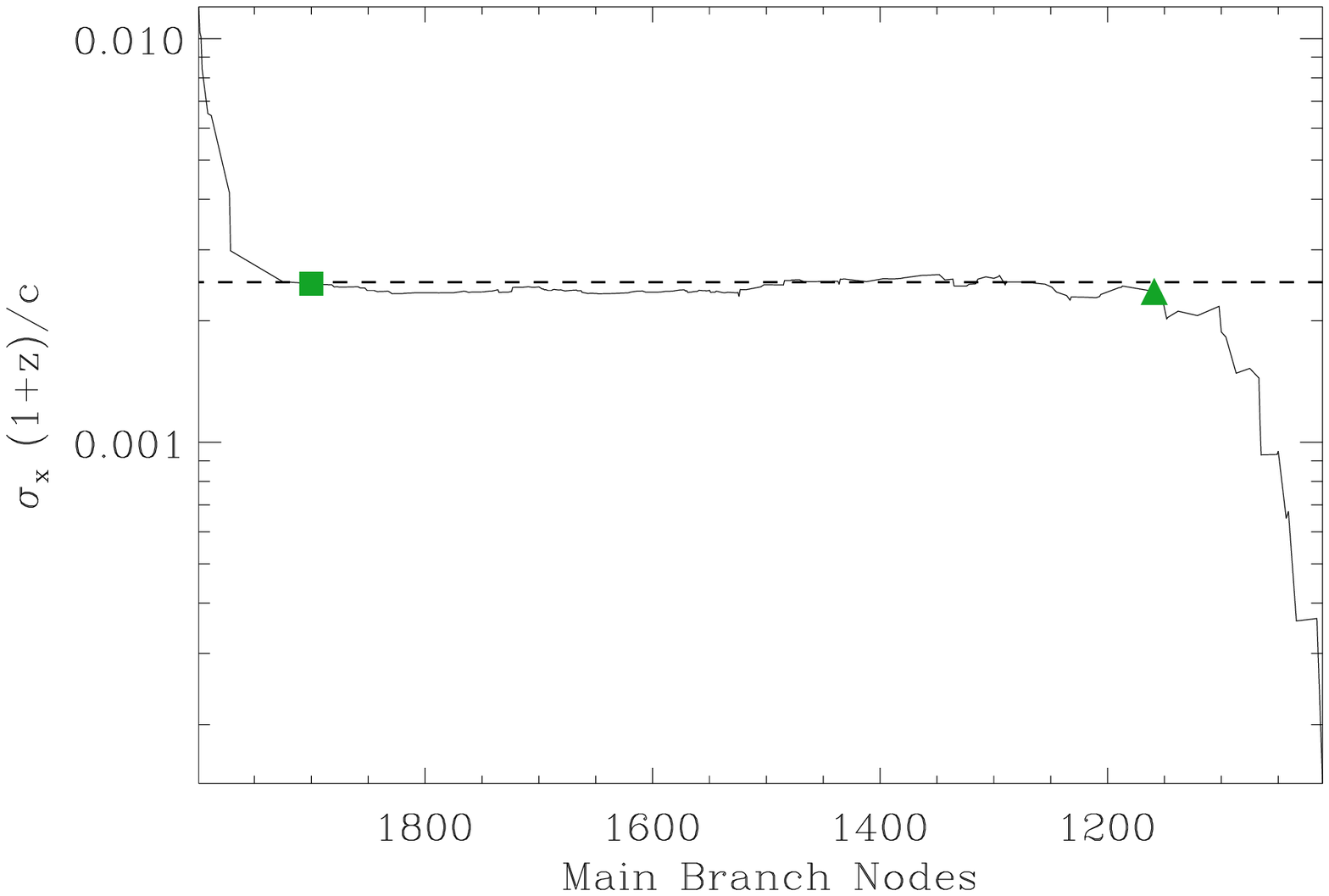}
\caption{Velocity dispersion of the leaves of each node
along the main branch of the binary tree shown in Figure \ref{fig:dendrogram}. The filled square and triangle indicate the final thresholds (nodes $x_1$ and $x_2$) chosen
by the algorithm. The dashed line shows the l.o.s. velocity 
dispersion of the particles within a 3D sphere of radius $3r_{200}$.}
\label{fig:thresholds}
\end{figure}

\begin{figure*}
\includegraphics[angle=0,scale=.70]{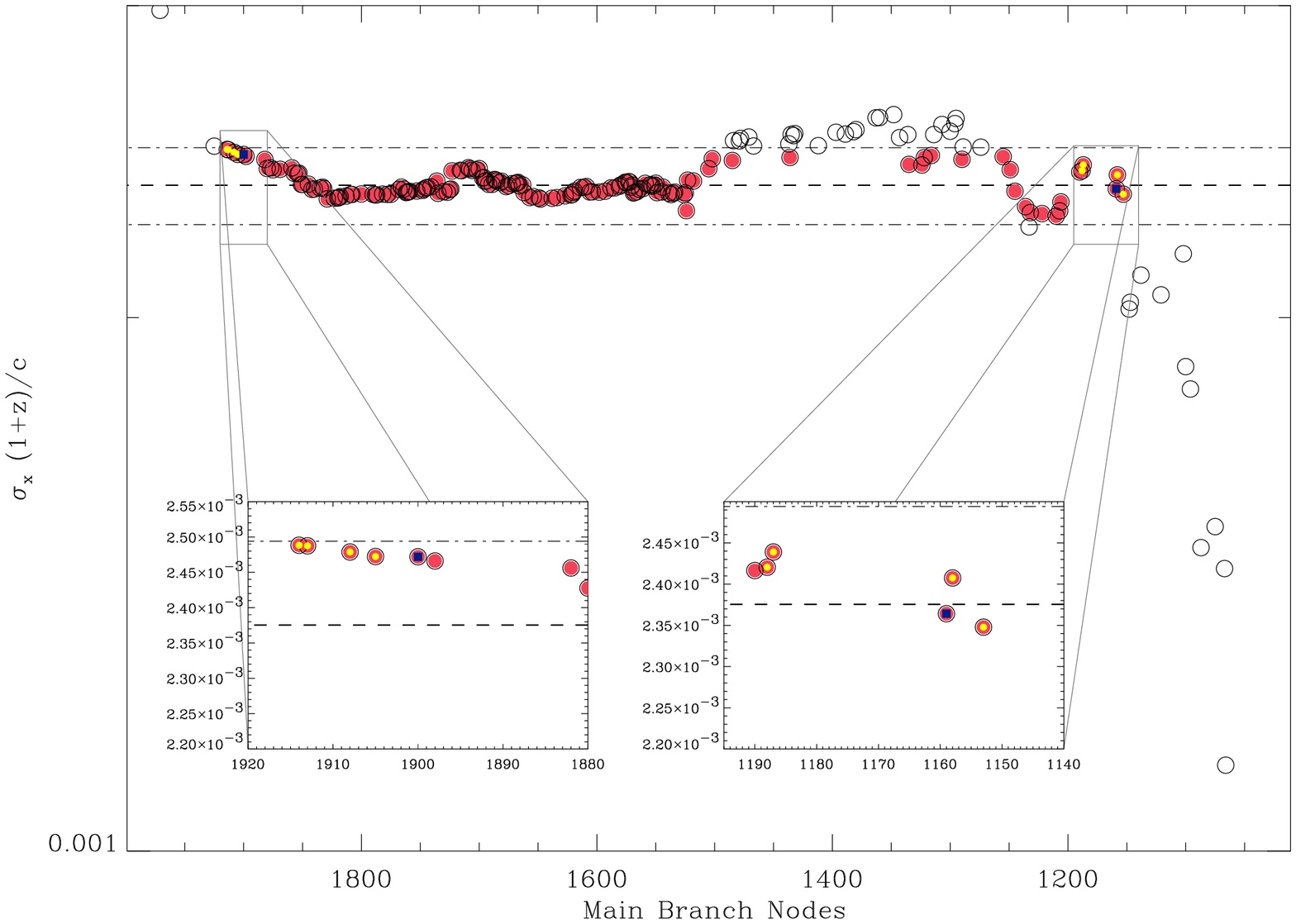}
\caption{Line-of-sight velocity dispersion of the leaves of each node along the main branch of 
the binary tree shown in Figure \ref{fig:dendrogram}. The filled 
red circles are the $N_\delta$ nodes. The dashed line indicates $\sigma_{\rm pl}$, whereas the dot-dashed lines show 
the range defined by equation (\ref{eq:sig-plateau}): $\delta=0.05$ in this case. 
The ten circles with a yellow centre indicate the five nodes closest to the root and the five nodes closest to the leaves. These nodes
also appear in the two insets. The two blue squares denote the two final nodes $x_1$ and $x_2$. }
\label{fig:plateau-def}
\end{figure*}

Gravity is a long-range force and defining a system of galaxies
is somewhat arbitrary. The most widely used approach is to select 
a system according to a density threshold, either in redshift
space, for real data, or in real space, for simulated data.

When applied to real data, the binary tree described
in the previous section has the advantage of building a hierarchy of galaxy
pairs with increasing, albeit projected, binding energy. The tree automatically
arranges the galaxies in potentially distinct groups based on a single 
parameter, the galaxy mass. To get effectively distinct groups, 
however, we need to choose a threshold, i.e. the node 
from which the group members hang, where we cut the tree.
This choice has been traditionally arbitrary (\citealt{materne78}; \citealt{serna96}).

D99 suggests a more objective criterion based on the following 
argument. We first identify the main branch of the binary tree, that is the branch
that links the nodes from which, at each level of the tree, the largest number of leaves hang. The velocity dispersion $\sigma_x$ of
the galaxies hanging from a given node $x$ shows a characteristic
behaviour when walking towards the leaves along
the main branch (Figure \ref{fig:thresholds}): initially it decreases rapidly, it then reaches
a long plateau and again drops rapidly towards the end of the
walk. The plateau is a clear indication of the presence of the
nearly isothermal cluster: at the beginning of the walk, $\sigma_x$ 
is large because of the presence
of background and foreground galaxies; at the end of the walk
the tree splits into different substructures and 
$\sigma_x$ drops again. 

The two nodes $x_1$ and $x_2$ that limit the $\sigma$ plateau are good candidates for the cluster/substructure identification. 
To locate $x_1$ and $x_2$, we proceed as follows.
We consider the density distribution of the $\sigma_x$. The mode  
of this density distribution is $\sigma_{\rm pl }$.
We then identify the $N_\delta$ nodes whose velocity dispersion $\sigma_x$ fulfills the inequality 
\begin{equation}
{\vert\sigma_{\rm pl}-\sigma_x\vert\over \sigma_{\rm pl}} \leqslant\delta \;.
\label{eq:sig-plateau}
\end{equation} 

Clearly, the number of nodes $N_\delta$ 
depends on the parameter $\delta$. We limit the parameter $\delta$ in the range $[0.03,0.1]$, but we
also compute the number of nodes $N_{0.3}$ when $\delta=0.3$. 
We determine the number of nodes $N_\delta$ corresponding to 
increasing values of $\delta\in [0.03,0.1]$ by step of 0.01 until
$N_\delta$ is larger than  $0.8N_{0.3}$. This sets the final number of nodes
$N_\delta$. If all the $N_\delta$ with $\delta\in [0.03,0.1]$ are smaller than
$0.8N_{0.3}$, we choose $0.8N_{0.3}$ as the final number of nodes.
The upper limit of the range of $\delta$, $\delta=0.1$, is 
chosen because it always provides a sufficiently
large number of nodes (15 in our sample); the arbitrary threshold $0.8N_{0.3}$ 
is chosen because it enables to deal efficiently with particularly peaked density distributions of $\sigma_x$. 
Among the $N_\delta$ nodes, we locate 
the five nodes closest to the root and the five nodes closest to the leaves; among the former set, 
the final node $x_1$ has $\sigma_x$ with the smallest discrepancy from $\sigma_{\rm pl}$, and
similarly for the final node $x_2$ among the five nodes closest to the leaves. This procedure,
illustrated in Figure \ref{fig:plateau-def}, guarantees a sufficiently large number 
of nodes along the $\sigma$ plateau and prevents us from  locating the extrema of the $\sigma$ plateau on 
nodes whose $\sigma_x$ are too discrepant from $\sigma_{\rm pl}$. 

\subsection{Redshift-space diagram, caustics and final members} \label{sec:redspace}

\begin{figure*}
\includegraphics[angle=0,scale=.52, bb=70 370 558 700]{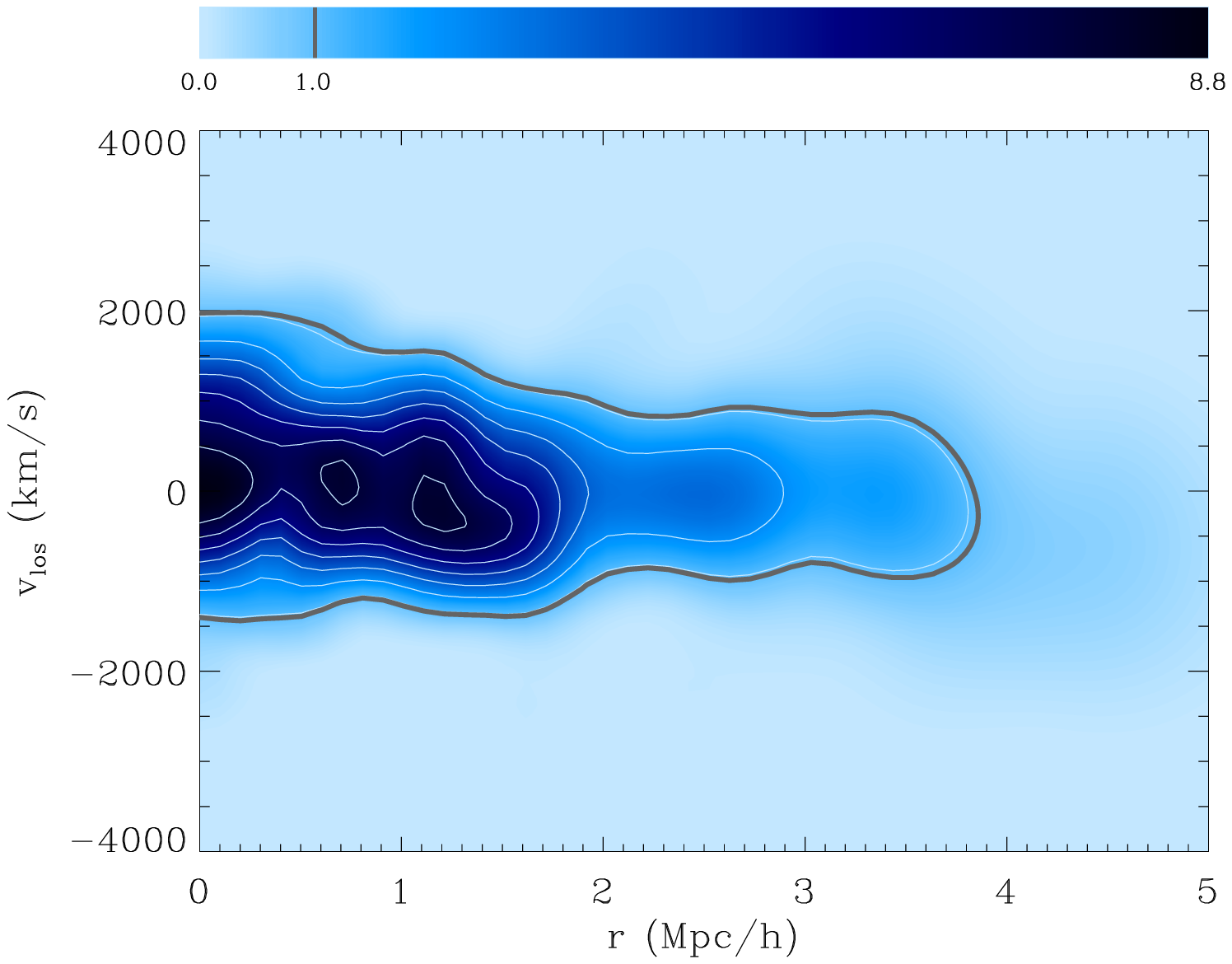}
\includegraphics[angle=0,scale=.45, bb=35 14 566 395]{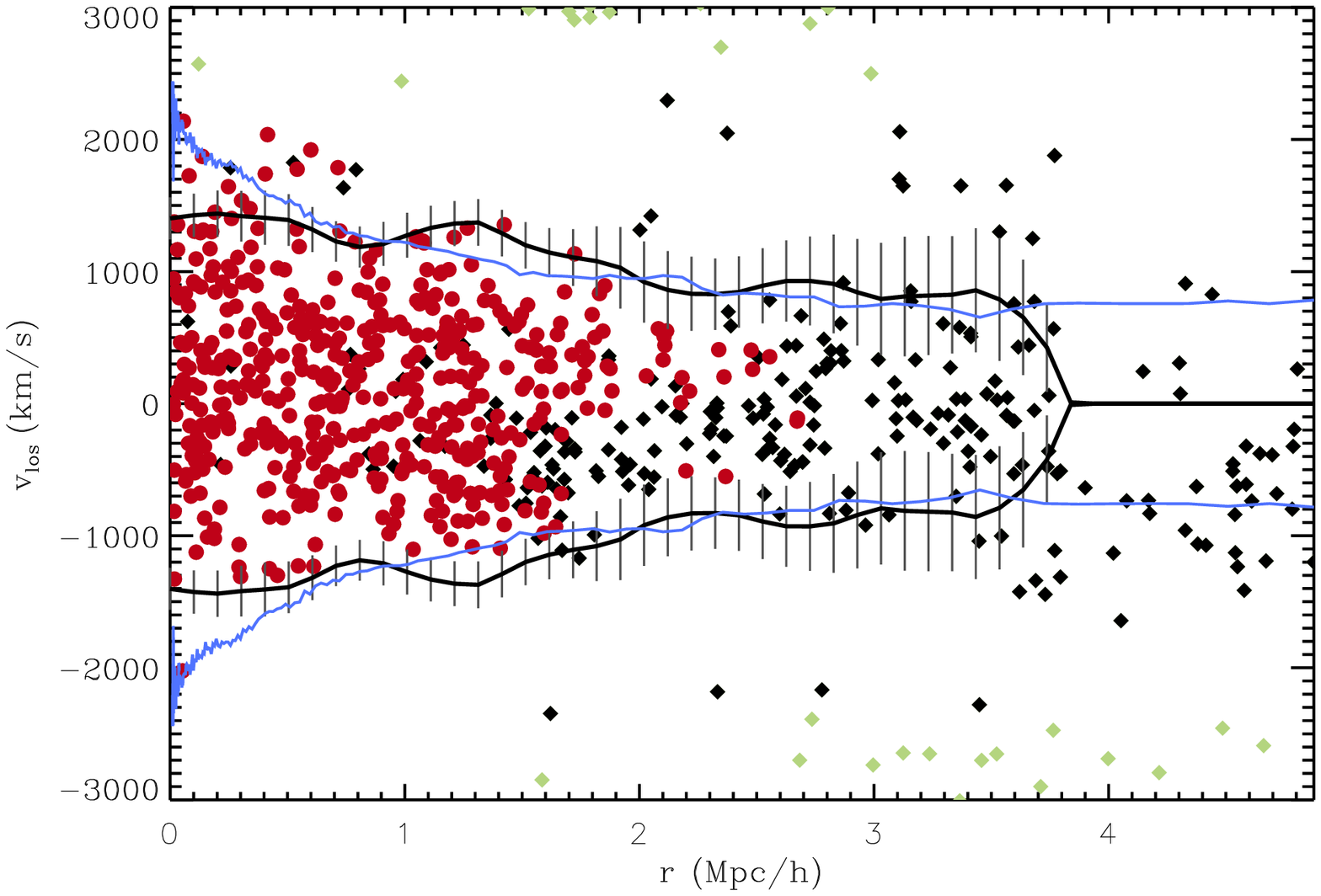}
\caption{Left panel: the function $f_q(r,v)$ on the redshift diagram (colour map and contours)
of the cluster shown in Figures \ref{fig:dendrogram}-\ref{fig:plateau-def}. 
The thick line shows where $f_q(r,v)=\kappa$.
Right panel: the redshift diagram of the same cluster with the upper and the lower caustics. The black and cyan  
lines are the estimated and true caustics respectively; 
the error bars are estimated with the procedure described in section \ref{sec:uncertainty}.
The symbols are the particles in the catalogue: the red dots are
the particles within a sphere of $3r_{200}$ centred on the cluster centre, in real space; the green diamonds are the particles whose l.o.s. velocity exceeds 3.5 times the velocity dispersion $\langle v^2\rangle^{1/2}$ 
of the candidate members.}
\label{fig:red-diag}
\end{figure*}

The candidate cluster members are the galaxies hanging from 
node $x_1$ and constitute the main group of the binary tree. These galaxies determine the centre and the size
$R$ of the cluster. 
The redshift $z_c$ of the cluster is the median of 
the candidate redshifts. The celestial coordinates of the centre are
the coordinates of the two-dimensional (2D) density peak of the candidates. To
find the peak, we compute the 2D density distribution $f_q(\alpha,\delta)$ 
of the candidates on the sky with the kernel method described
below (equation \ref{MM1}). In general, the smoothing parameter $h_c$ is
automatically chosen by the adaptive kernel method. However, 
to save substantial computing time, here we set $h_c=0.15 (D_A/320 h^{-1} {\rm Mpc})$~rad, 
where $D_A(z_c)$ is the angular diameter distance of the cluster. This choice yields accurate results.
The cluster size $R$ is the mean projected separation of the candidate cluster members
from the cluster centre. 

We then build the redshift diagram, which is the plane $(r,v)$ of the
projected distance $r$ and the l.o.s.
velocity $v$ of the galaxies from the cluster centre.
If $\psi$ is the angular separation between the cluster
centre and a galaxy at redshift $z$, then 
\begin{equation}
 r = {cD_A(z_c) \over H_0} \sin\psi \; ,
\end{equation}
and 
\begin{equation}
 v = c{z-z_c\over 1+z_c} \; ,
\end{equation}
where we have assumed that the galaxy velocity within the 
cluster is much smaller than the speed of light $c$ and we have neglected the peculiar
velocity of the cluster.
Note that to avoid artificial depletion of the caustic amplitude
at small $r$ due to the small number of galaxies in the central
region, the galaxy distribution is mirrored to negative 
$r$. As an example, in Figure \ref{fig:red-diag} we show the redshift diagram of 
the system shown in Figures \ref{fig:dendrogram}-\ref{fig:plateau-def}. 

It is now necessary to locate the caustics. The distribution 
of $N$ galaxies in the redshift diagram is described by the 2D distribution:

\begin{equation}
f_q({\bf x})={1\over N}\sum_{i=1}^{N} {1\over h_i^2} K\left({\bf x}-{\bf x}_i\over h_i\right)
\label {MM1}
\end{equation}  
where ${\bf x}=(r,v)$, the adaptive kernel is 
\begin{equation}
K({\bf t})=\cases { 4\pi^{-1} (1- t^2)^3 & $ t <1$\cr
	0 & otherwise; \cr}
\label {MM2}
\end{equation}
$h_i=h_ch_{\rm opt}\lambda_i$ is a local smoothing length, 
$\lambda_i=[\gamma/f_1({\mathbf x}_i)]^{1/2}$, $f_1$ is equation (\ref{MM1}) where $h_c=\lambda_i=1$
for any $i$, and $\log\gamma = \sum_i\log[f_1({\mathbf x}_i)]/N$. D99 describes 
how the adaptive kernel method automatically determines $h_c$.

The optimal smoothing length
\begin{equation}
h_{\rm opt} = {6.24 \over N^{1/6}} \left(\sigma_r^2+\sigma_v^2\over 2\right)^{1/2} 
\label {MM3}
\end{equation}
depends on the marginal standard deviations
$\sigma_r$ and $\sigma_v$ of the galaxy coordinates in the $(r,v)$ plane. These two $\sigma$'s must have
the same units and we require the smoothing to mirror the uncertainty in the 
determination of the position and velocity of the galaxies. We therefore 
rescale the coordinates such that $q=\sigma_v/\sigma_r$ assumes a
fixed value. We choose $q=25$. With this value, 
an uncertainty of 100 km s$^{-1}$ in the $v$ direction, for example, 
weights like an uncertainty of $0.02 h^{-1}$ Mpc in the $r$ direction. 
The optimal smoothing length $h_{\rm opt}$ adopted by D99
is half the value we use here. However, that $h_{\rm opt}$ is
derived in the context of the probability density estimate 
under the assumption that ${\bf x}=(r,v)$ is a normally
distributed random variate with unit variance 
\citep[sects. 4.3 and 5.3]{silverman86}. However, this assumption clearly does not apply to our context here, and we find that
we obtain more accurate results with equation (\ref{MM3}).

The caustics are now the 
{\it loci} of the pairs $(r,v)$ that satisfy the caustic equation
\begin{equation}
f_q(r,v)=\kappa \; , 
\label{eq:caustic}
\end{equation}
where $\kappa$ is a parameter we determine below. 
In general, the density distribution $f_q(r,v)$ is not symmetric around the straight line $v=0$ 
 in the $(r,v)$ plane,
and equation (\ref{eq:caustic}) determines two curves $v_{\rm up}(r)$ and $v_{\rm down}(r)$,
above and below this line. 
Because we assume spherical symmetry, we define the two caustics to be
${\cal V}_\kappa^\pm(r) = \pm {\rm min}\{\vert v_{\rm up}(r)\vert, \vert v_{\rm down}(r)\vert\}$.
This choice limits the number of interlopers.
The caustic amplitude is  ${\cal A}_\kappa(r)=[{\cal V}_\kappa^+(r)-{\cal V}_\kappa^-(r)]/2$.
Any realistic models of galaxy clusters has ${\rm d}\ln {\cal A}/{\rm d}\ln r\la 1/4$.
A value of this derivative much larger than $1/4$ indicates
that the algorithm has found an incorrect location of the caustics, which includes an excessive number of foreground and/or
background galaxies. Therefore, whenever ${\rm d}\ln {\cal A}/{\rm d}\ln r > \zeta$ at a given $r$, the algorithm
replaces ${\cal A}(r)$ with a new value that yields ${\rm d}\ln {\cal A}/{\rm d}\ln r =1/4$. We choose $\zeta =2$,
rather than  $\zeta =1/4$, to keep this constrain loose.

We choose the parameter $\kappa$, that determines the
correct caustic location, as the root of the equation 
\begin{equation}
S(\kappa)\equiv\langle v_{\rm esc}^2\rangle_{\kappa,R}-4\langle v^2\rangle = 0 \; ,
\label{eq:skappa}
\end{equation}
where $\langle v_{\rm esc}^2\rangle_{\kappa,R}=\int_0^R{\cal A}_\kappa^2(r)\varphi(r){\rm d}r/
\int_0^R\varphi(r){\rm d}r$ is the mean caustic amplitude within the main group size $R$,
$\varphi(r)=\int f_q(r,v) {\rm d}v$, and $\langle v^2\rangle^{1/2}$ is
the velocity dispersion of the candidate cluster members with respect
to the median redshift.
When $S(\kappa)=0$, the escape velocity inferred from the caustic amplitude
equals the escape velocity of a system in dynamical equilibrium with a Maxwellian 
velocity distribution within $R$. However, it is important to emphasize that the prescription for determining $\kappa$ 
works equally well for any cluster independently of its dynamical status within $R$. Therefore 
this prescription appears to be just a convenient recipe to locate
the caustics properly and does not limit the caustic technique to systems 
in equilibrium within $R$. 

The outcome of the process described in the preceding paragraphs is outlined in Figure \ref{fig:red-diag}, where 
${\cal V}_\kappa^\pm(r)$ are shown together with the expected caustic amplitude
$v_{\rm esc, los}(r) = \pm \sqrt{2\mid\phi(r)\mid/g(\beta)}\equiv \pm \sqrt{\phi_\beta(r)}$. Figure \ref{fig:red-diag} 
shows a good agreement between the expected and estimated caustics; in Sect. \ref{sec:GP}, we
will see that this is a general result.

\subsection{Tunable parameters}

The caustic technique requires two sets of parameters: a set for building the binary
tree and identifying the cluster candidate members and a set 
for the location of the caustics in the redshift diagram. Most parameters
are automatically determined by the method with the prescriptions described above.
A few remaining parameters,
listed in Table \ref{tab:params}, are chosen in input. In most cases, keeping these parameters to the default value
returns accurate results. In fact, in Section \ref{sec:finetuning} we show how, in general, the improvement 
provided by a fine tuning of these parameters is modest. 

\begin{table}
\begin{tabular}{lcc}
\hline
\hline
parameter & symbol    &    default value \\
\hline
\hline
\multicolumn{3}{c}{Binary tree construction and candidate members} \\
galaxy mass              &  $m$    &  $10^{12}h^{-1}{\rm M}_\odot$      \\
cluster threshold node              &  $x_1$   &  $-$         \\
substructure threshold node              &  $x_2$   &  $-$      \\
\hline
\multicolumn{3}{c}{Caustic location} \\
rescaling              &  $q$    &  25       \\
smoothing length             &  $h_c$   &  $-$        \\
threshold             &  $\kappa$   &  $-$      \\
derivative limit of ${\cal A}$        &  $\zeta$   &  2       \\
\hline
\multicolumn{3}{c}{Mass profile} \\
filling factor             & ${\cal F}_\beta$   &  $0.7$          \\
\end{tabular}
\caption{List of the tunable parameters.}
\label{tab:params}
\end{table}

\section{The caustic technique at work}

\begin{figure*}
\includegraphics[angle=0,scale=.63]{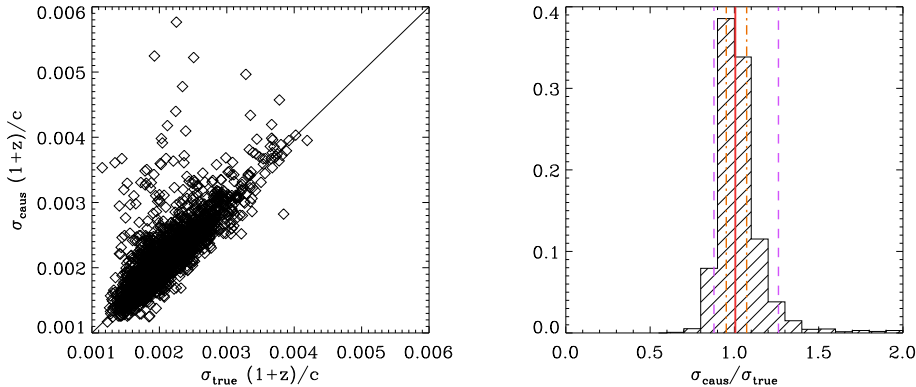}
\caption{Comparison between the true velocity dispersion $\sigma_{\rm true}$ and the velocity dispersion 
of the cluster candidate members $\sigma_{\rm caus}$. The red solid line in the histogram in the right panel represents the median and the dash-dotted orange (dashed violet) lines represent the 50 (90) percent range.}
\label{fig:sigmacaustrue}
\end{figure*}

We now apply the caustic method to the mock redshift catalogues
described in Sect. \ref{subsec:mockcatalogue}. We describe
how the technique recovers the centre and velocity dispersion
of a cluster, and its escape velocity and mass profiles.
Finally, we provide a simple recipe to estimate the 
uncertainties on these profiles.
 
\subsection{Global properties of the cluster}\label{sec:CGP}

In order to remove foreground and background large-scale structure from the redshift diagram,
we remove all the particles with  l.o.s. velocity larger than 3.5 times the velocity dispersion
of the candidate members. Nevertheless, when the cluster
is embedded in a particularly
crowded area of the sky, it can  also happen that 
the main branch of the tree identifies a cluster different
from the target cluster. These cases can happen, for example, when
the target cluster is less rich than nearby systems;
they are easily solved by limiting the input particle 
catalogue to a small enough region centred on the target cluster.  
In our sample, the main branch of the binary tree identifies a cluster
different from our target 326 times out of 3000, in other words only 11 percent of the times. 
For the sake of simplicity, 
we limit our sample to the 2674 clusters 
which are identified without further limitation of the cluster field of view. 

The input centre $[\alpha, \delta, cz]=[90^\circ, 0^\circ, 32000\, {\rm km\, s}^{-1}]$, is recovered very well: 
the percentile ranges are $\alpha=[89.96, 90.00, 90.04]^\circ$,
$\delta = [-0.040, 0.000, 0.036]^\circ$, and
 $cz=[31740, 31998, 32222]$ km s$^{-1}$.

Figure \ref{fig:sigmacaustrue} shows the agreement between the estimated velocity dispersion $\sigma_{\rm caus}$ (the velocity dispersion of the galaxies hanging from the node $x_1$) and the true velocity dispersion $\sigma_{\rm true}$, 
defined as the l.o.s. velocity dispersion of the particles within a sphere of radius $3r_{200}$ in real space:
in 50 (95) percent of the systems the estimated velocity dispersion is within 5 (30) percent of
the real one.

\subsection{The escape velocity profile}\label{sec:GP}

\begin{figure}
\includegraphics[angle=0,scale=.62, bb=70 10 504 330]{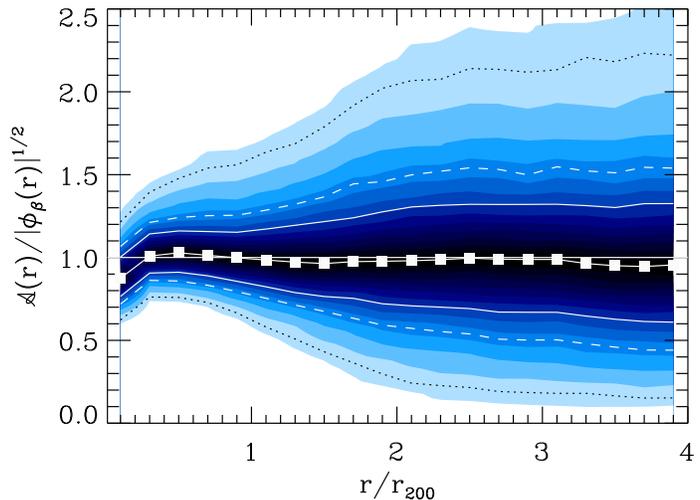}
\caption{Profiles of the ratio between the caustic amplitude ${\cal A}(r)$ and 
the l.o.s. component of the true escape velocity $\langle v_{\rm esc, los}^2(r)\rangle 
= -2\phi(r)/g(\beta)\equiv \phi_\beta(r)$.
The numerical gravitational potential profile $\phi(r)$ is
derived from the true mass distribution within $r_{\rm max}=10 h^{-1}$ Mpc from
the cluster centre: 50, 68, and 90 percent of the profiles are within the upper and lower solid, dashed, and dotted curves,
respectively. The solid squares show the median profile. The darkness of the shaded areas is proportional
to the profile number density on the vertical axis. }
\label{fig:caus-ratio}
\end{figure}

The caustic amplitude is related to $\phi_\beta(r)$ (equation \ref{eq:rig-pot}), the mean component along the 
line of sight of the escape velocity, which is a combination of
the gravitational potential and the velocity anisotropy parameter. 
Figure \ref{fig:caus-ratio} shows how well the caustic amplitude recovers
$\phi_\beta(r)$. On average, the potential is slightly underestimated in the central region ($r \lesssim 0.2\,r_{200}$), but in the outer regions the agreement with the true escape velocity is remarkable. 
The slight systematic underestimate at the centre is due
to the small number of particles at very small radii in the redshift diagram, that leads
to an underestimate of the caustic amplitude.
In the cluster outskirts, the location of the caustics 
is less accurate because the number of particles sampling the density distribution
$f_q(r,v)$ is smaller. This fact increases the scatter of the profiles  at
large radii. We will see below that increasing the number of particles in the
mock catalogues indeed reduces the scatter (Figure \ref{fig:caus-ratio-2000}). We remind, however,
that the statistical significance of the scatter decreases at increasing radii: in fact, 
the number of clusters contributing to the median profile decreases at large $r$, 
because we exclude those clusters that have a null caustic amplitude ${\cal A}(r)=0$
beyond any given radius $r$. 

\subsection{The mass profile}\label{sec:mass}

\begin{figure}
\includegraphics[angle=0,scale=.70,bb=20 14 340 280]{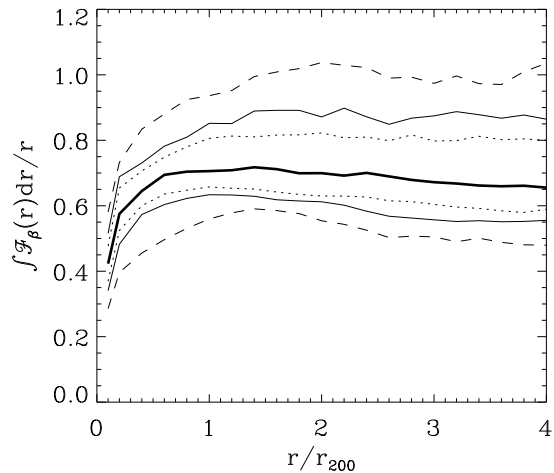}
\caption{Profiles of the integral $\int_0^r{\cal F}_\beta(x) {\rm d}x/r$ described
in the text; 90, 68, and 50 percent of the profiles are within the upper and lower dashed, solid and dotted curves.
The central solid curve is the median profile.}
\label{fig:fbeta-int}
\end{figure}

DG97 show that the caustic amplitude can be related to the cumulative mass
profile of the cluster by the relation (\ref{eq:rig-massprof}):
\begin{displaymath}
GM(<r)=\int_0^r {\cal F}_\beta(r) {\cal A}^2(r) dr \; .
\end{displaymath}
The bottom right panel of Figure \ref{fig:beta-4pan} shows the function ${\cal F}_\beta(r)$ in our simulations. 
At radii in the range $\sim (0.5-4) r_{200}$, the average ${\cal F}_\beta(r)$ has
a mild variation, between 0.5 and 0.8. This result led DG97 and D99 to 
assume ${\cal F}_\beta(r)={\rm const}$ {\it tout court} and
assume that the mass profiles of real clusters can be estimated with the expression (\ref{eq:recipe-massprof}):
\begin{displaymath}
GM(<r)={\cal F}_\beta \int_0^r {\cal A}^2(r) dr \; .
\end{displaymath}
We can choose the correct value of the factor ${\cal F}_\beta$  
by considering the contribution of the filling function ${\cal F}_\beta(r) $
in the integral of equation (\ref{eq:rig-massprof}). Figure \ref{fig:fbeta-int}
shows $\langle {\cal F}_\beta(r)\rangle = \int_0^r {\cal F}_\beta(x){\rm d}x/r$, where ${\cal F}_\beta(x)$ 
is the profile of each individual cluster. 
At radii larger than $\sim 0.5 r_{200}$, $\langle{\cal F}_\beta(r)\rangle$
is basically constant and supports the validity of equation (\ref{eq:recipe-massprof}). 

We see that the most appropriate value is ${\cal F}_\beta=0.7$.
This choice disagrees with the value ${\cal F}_\beta=0.5$ adopted by DG97 and D99.
In this early work, the algorithm for the determination of the $\sigma$ plateau 
was less accurate than our algorithm here and systematically provided slightly larger caustic amplitudes.
This overestimate was compensated
by a smaller ${\cal F}_\beta$, that, on turn, returned the correct
mass profile, on average. Here, our improved algorithm 
 appears to be more appropriate because
it returns the correct $\phi_\beta(r)$ profile (Figure \ref{fig:caus-ratio}) and,
in order to estimate the correct mass profile, requires
a value of ${\cal F}_\beta$ in agreement with what can be
expected by inspecting Figure  \ref{fig:fbeta-int}. 

\begin{figure}
\includegraphics[angle=0,scale=.62, bb=70 10 504 330]{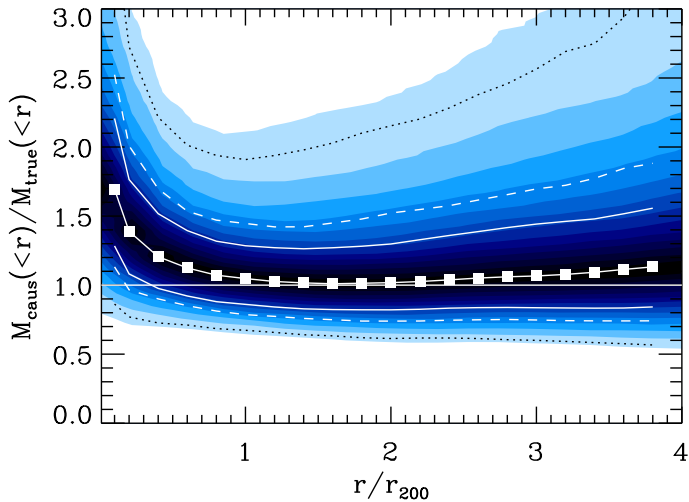}
\caption{Profiles of the ratio between the caustic and the true mass profile, adopting equation 
(\ref{eq:recipe-massprof}) and ${\cal F}_\beta=0.7$.
The lines and shaded areas are as in Figure \ref{fig:caus-ratio}.}
\label{fig:mass-newF}
\end{figure}

Figure \ref{fig:mass-newF} shows that, on average, the mass profile is 
estimated at better than 10 percent at radii larger than $\sim 0.6 r_{200}$. 
Clearly, at smaller radii, the assumption ${\cal F}_\beta(r)={\rm const}$
breaks down and the mass 
is severely overestimated. 

As already suggested by DG97, if we assume that the cluster
is in virial equilibrium in the central region, we can
use the virial theorem to estimate the mass there and 
limit the use of the caustic method to the cluster outskirts alone, 
where the equilibrium assumption does not hold. 
Here we use the virial theorem and the median and average mass
estimators from \citet{hei85} to estimate the mass within $\alpha R$, where
$R$ is the mean clustrocentric separation of the candidate cluster members
from the binary tree (see Section \ref{sec:redspace}) and $\alpha$ is a free
parameter. In our sample, the percentile
range is $R=[0.50,1.23,1.68] h^{-1}$~Mpc. 
We compute the ratio between the estimated mass and the true mass for
different values of $\alpha$. We find that 
the best estimates are obtained when $\alpha=0.7$. In this case, the ratio
between the estimated and true masses is, on average, 
$1.03$ for the virial theorem, 1.30 and 1.49 for the median and average
mass estimators, respectively. Different values of $\alpha$ yield worse
mass estimates. This result indicates that the radius $0.7 R$ generally contains 
the cluster region in approximate virial equilibrium.

\begin{figure}
\includegraphics[angle=0,scale=.61,bb=65 10 504 320]{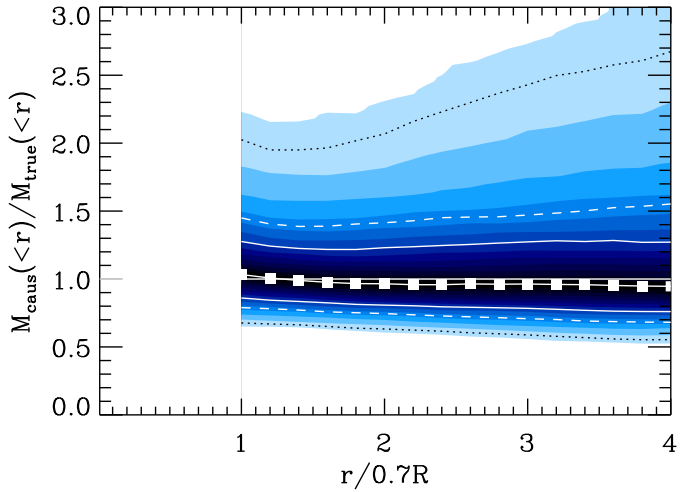}
\caption{Profiles of the ratio between the caustic and the true mass profile, when combined with the virial masses. 
The lines and shaded areas are as in Figure \ref{fig:caus-ratio}. }
\label{fig:mass-vir}
\end{figure}

When we estimate the mass with the virial theorem within $0.7R$ and
use equation (\ref{eq:recipe-massprof}) with $0.7R$ as the lower limit
of the integral, we still obtain a very good estimate of the real mass
(Figure \ref{fig:mass-vir}).

\subsection{The gravitational potential profile}

\begin{figure*}
\includegraphics[angle=0,scale=.78]{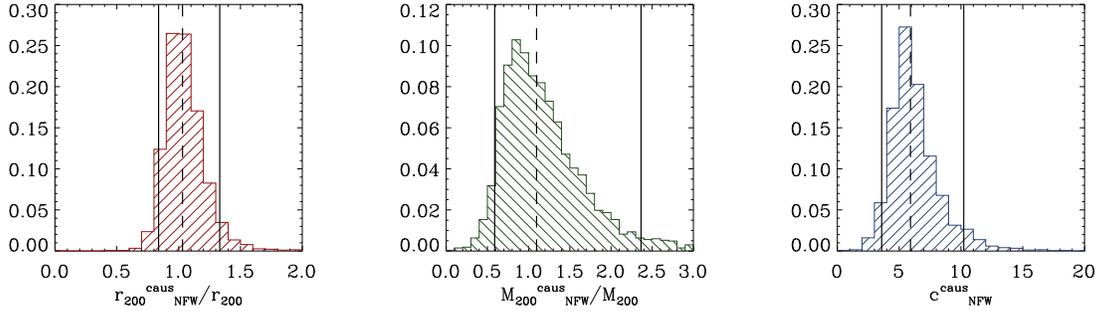}
\caption{Distribution of the ratio between the parameters of the best NFW fit to the caustic mass
profile and the true $M_{200}$ and $r_{200}$ (left and central panels) and distribution
of the concentration parameter of the best NFW fit to the caustic mass
profile (right panel). The dashed lines indicate the median
while the solid lines show the 90 percent range.}
\label{fig:nfw-ratio}
\end{figure*}

Within $r=1 h^{-1}$ Mpc, we fit the caustic mass profile of each 
individual cluster with an NFW profile.
The scale factor $r_{200}$ derived from this fit agrees
with the true $r_{200}$ derived in 3D;  the percentile range of their ratio is $[0.84, 1.03, 1.33]$.
 Consequently, for $M_{200}$, which is proportional to $r_{200}^3$, 
we have $[0.59,1.10,2.36]$ (Figure \ref{fig:nfw-ratio}). 

However, the concentration parameter $c$ derived from the NFW fit to
the caustic mass profile tends to overestimate by 36 percent, on average, the concentration parameter
of the NFW fit to the 3D mass profile: in fact, the caustic $c$ has percentile range $[3.59,5.92,10.23]$,
whereas we find $[2.06,4.62, 6.50]$ in 3D;
the ratio of these two parameters
has percentile range $[0.73, 1.36, 3.26]$.
This result is an obvious consequence of the mass overestimate at small radii (Figure \ref{fig:mass-newF}).

\begin{figure*}
\includegraphics[angle=0,scale=.52, bb=50 10 504 325]{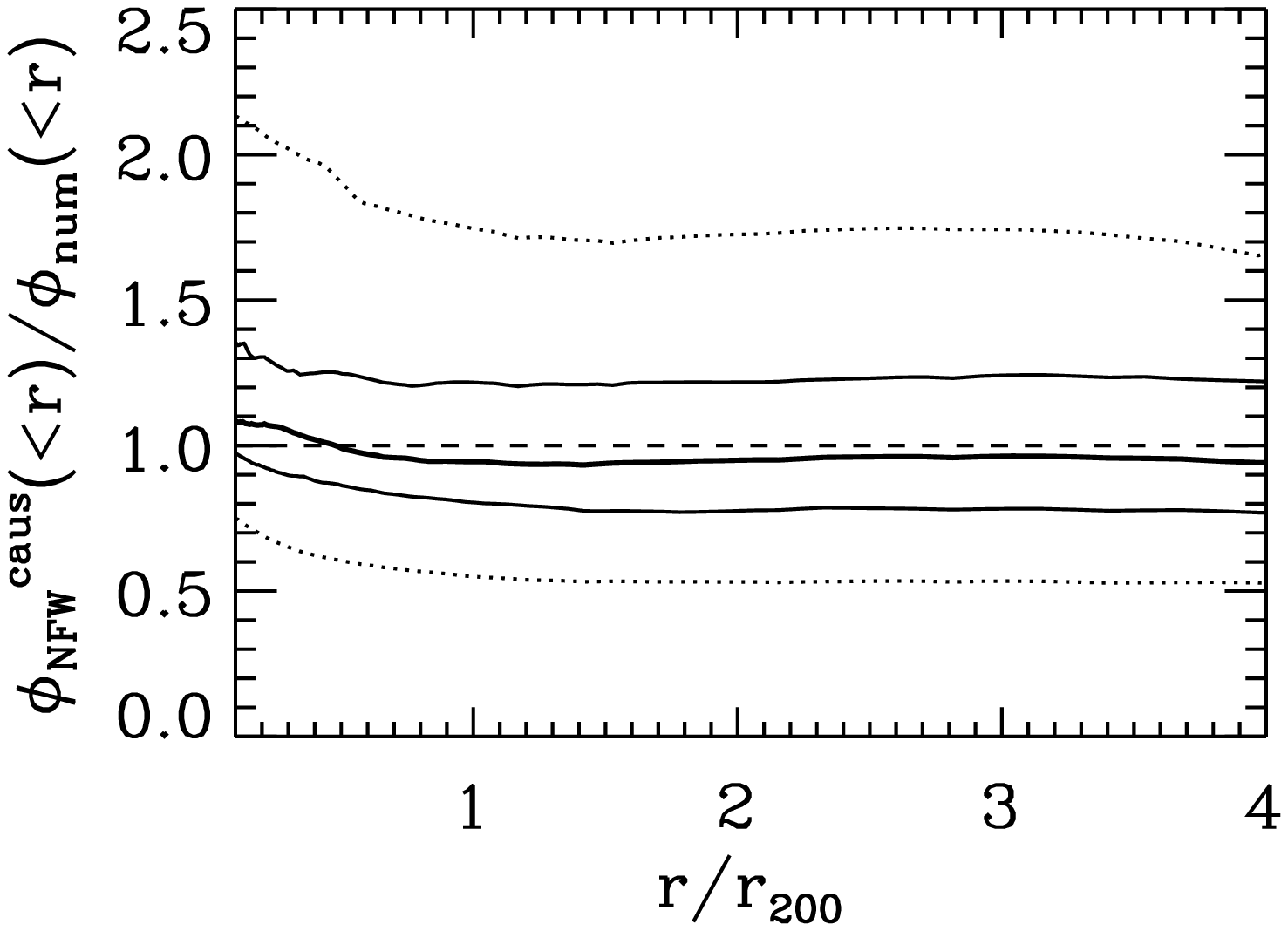}
\includegraphics[angle=0,scale=.52, bb=50 10 504 325]{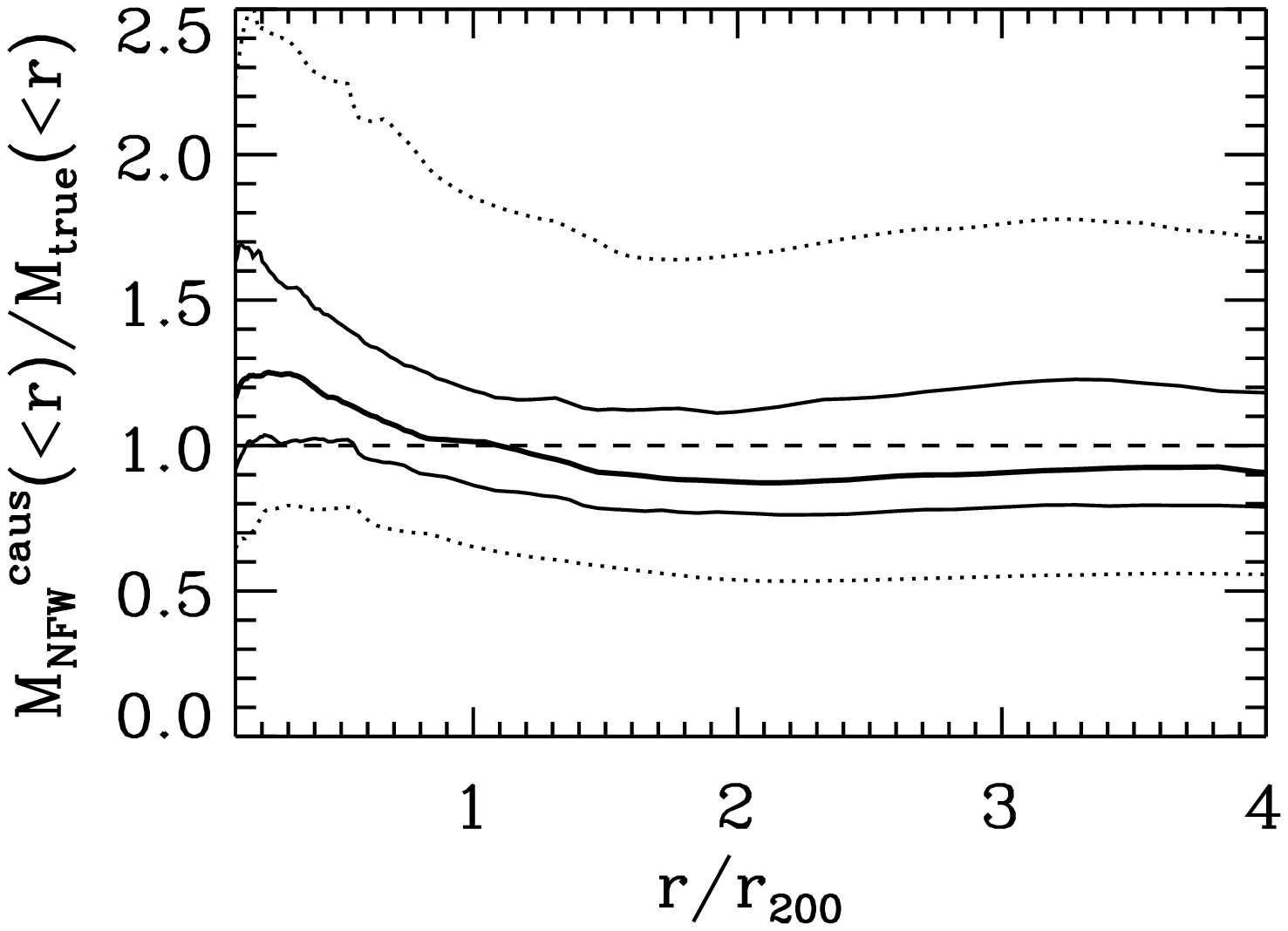}
\caption{Left panel: Profiles of the ratio between the NFW gravitational potential profile, with parameters derived
from the NFW best fit to the caustic mass profile, and the numerical gravitational potential profile 
derived from the true mass distribution within $r_{\rm max}=10 h^{-1}$ Mpc from
the cluster centre. Right panel: Profiles of the ratio between the 
NFW best fit to the caustic mass profile, and the true mass profile.
In both panels,  90 and 50 percent of the profiles are within the upper and lower dotted and solid curves,
respectively.
The central solid curves are the median profiles. For each cluster, the NFW fit to the mass profile is only
performed to the mass distribution within $1 h^{-1}$ Mpc.}
\label{fig:mass-nfw-ratio}
\end{figure*}

The NFW fit parameters can be used to derive the gravitational potential profile of the cluster.
The left panel of Figure \ref{fig:mass-nfw-ratio} shows that the estimated gravitational potential profile
returns the true profile within 10 percent, on average, and, in 50 percent of the cases, the estimated
profile is within 25 percent from the real one out to $4r_{200}$. This result is impressive
and derives from the mild radial variation of the functions ${\cal F}(r)$ and $g(\beta)$
in hierarchical clustering models.
 
The right panel of Figure \ref{fig:mass-nfw-ratio} shows the ratio between the caustic mass profile
derived from the NFW fit and the true mass profile. The agreement is again good,
within 20 percent, although not as good as in Figure \ref{fig:mass-newF}, because the caustic estimate 
is biased low due to the overestimate of $c$.

\subsection{Uncertainty estimate}
\label{sec:uncertainty}

\begin{figure*}
\includegraphics[angle=0,scale=.52, bb=70 10 504 330]{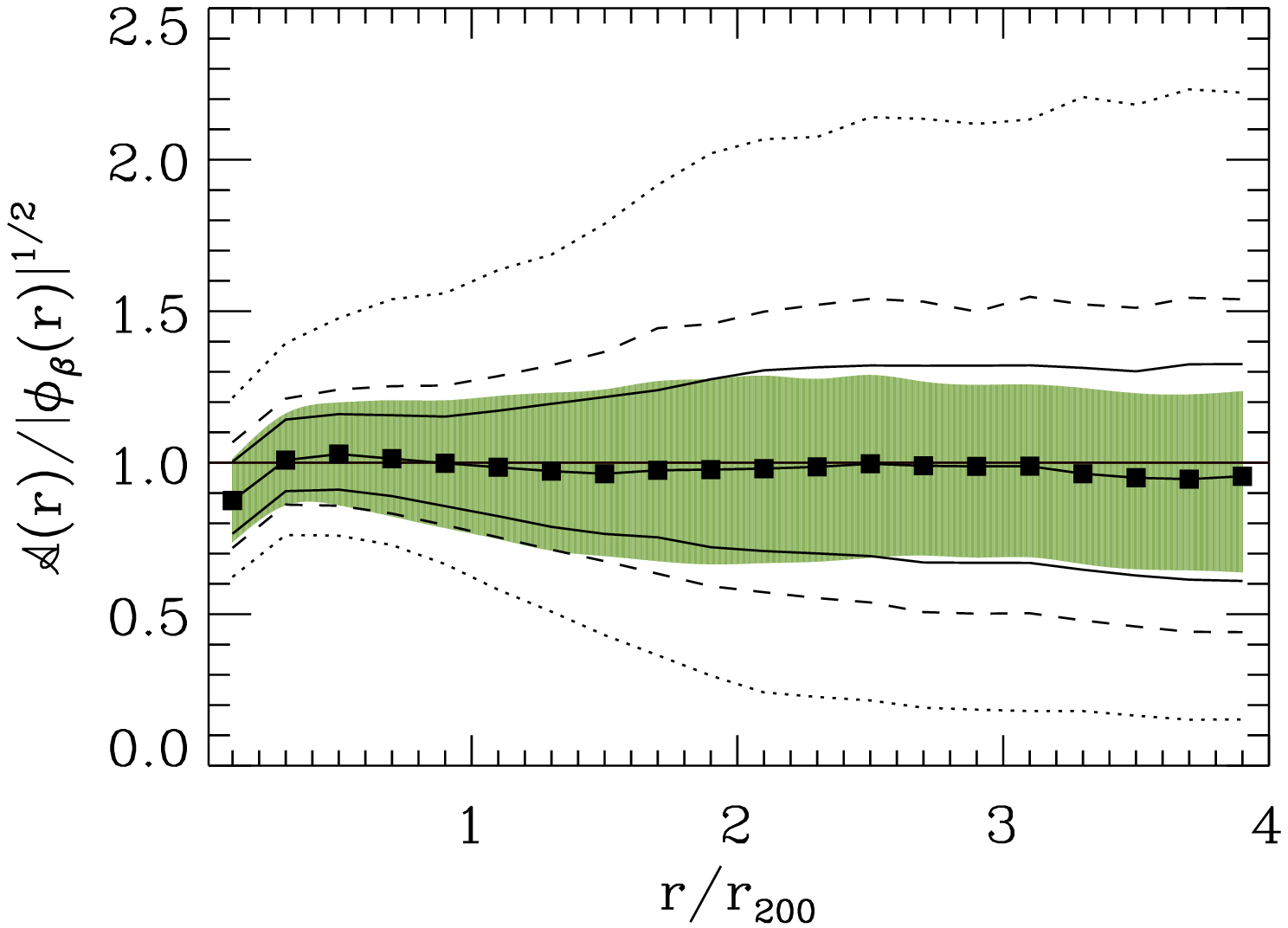}
\includegraphics[angle=0,scale=.52, bb=70 10 504 330]{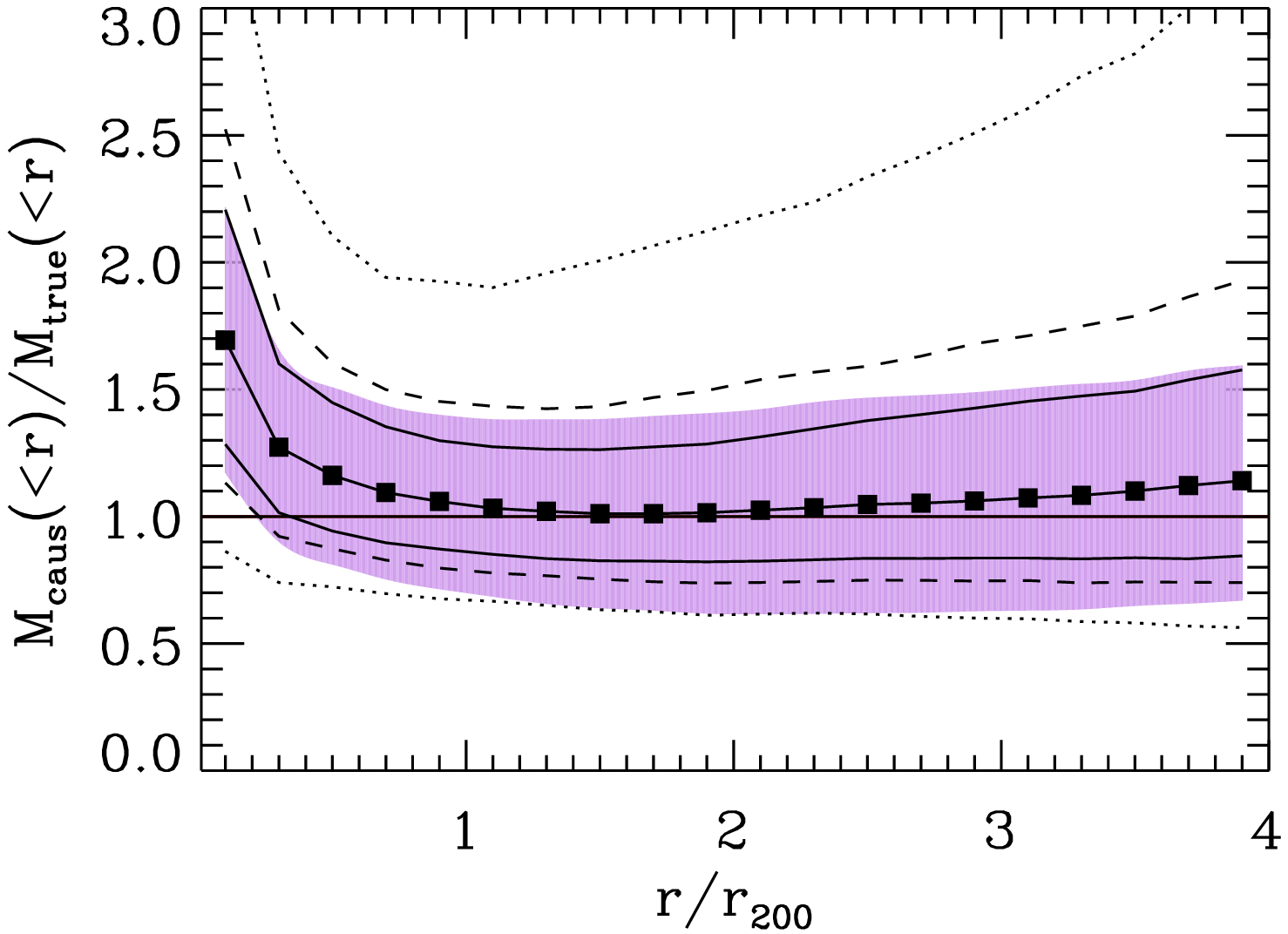}
\caption{The shaded areas show the median uncertainties estimated from equations (\ref{eq:errpot}) and (\ref{eq:errmass}). 
For comparison, we also show the 90 (dotted lines), 68 (dashed lines), and 50 (solid lines) percent ranges of the profiles from Figure \ref{fig:caus-ratio}.
The solid squares show the median profiles.}
\label{fig:errors}
\end{figure*}

Figures \ref{fig:caus-ratio} and \ref{fig:mass-newF} show the spread 
derived from the distribution of the profiles of the individual clusters. This spread is 
mostly due to projection effects, as we will see below (Sect. \ref{sec:projeff}). 
We remind here the simple recipe suggested in D99 to estimate this spread
when observing a real cluster of galaxies. 

The uncertainty in the measured value of ${\cal A}(r)$ depends on the
number of galaxies contributing to the determination of ${\cal A}(r)$;
we thus define the relative error 
\begin{equation}
\delta{\cal A}(r)/{\cal A}(r)=\kappa/\max\{f_q(r,v)\} , 
\label{eq:errpot}
\end{equation}
where the maximum value of $f_q(r,v)$  is 
along the $v$-axis at fixed $r$. The resulting error on the cumulative mass profile is
\begin{equation}
\delta M_i=\sum_{j=1,i}\vert 2m_j\delta{\cal A}(r_j)/{\cal A}(r_j)\vert , 
\label{eq:errmass}
\end{equation}
where $m_j$ is the mass of the shell $[r_{j-1},r_j]$ and $i$ is the 
index of the most external shell. 

The shaded bands in Figure \ref{fig:errors} are the median spreads
derived from the above equations. 
The median uncertainty on the escape velocity profile increases up to $r \sim 1.5 r_{200}$
and remains within $\sim 20-30$ percent at larger radii.
The mass profile has a slightly larger uncertainties but remains below $\sim 30-40$ percent out
to $r\sim 3r_{200}$. At radii smaller than $\sim 0.6r_{200}$ the assumption of
a constant filling function ${\cal F}_\beta(r)$ breaks down, as we mentioned above, but
this mass overestimate is systematic and can be easily taken into account.

Figure \ref{fig:errors} shows that our recipes for the estimate of
the uncertainties yield values in good agreement with the 50 percent range of 
the profiles.
Therefore, when applied to real clusters, this 
algorithm provides uncertainties with 50 percent confidence levels.

\section{Systematics}\label{sec:system}

In this section we investigate the systematic
errors affecting the estimate of the escape velocity and
mass profiles: the systematic errors can originate from 
the centre location, the parameters of the algorithm, and
the projection effects. We also show how the accuracy of the 
estimate depends on the number of objects in the
redshift catalogue.

\subsection{Centre identification}\label{clusid}

\begin{figure}
\hspace{-1cm}
\includegraphics[angle=0,scale=.62, bb=20 10 504 300]{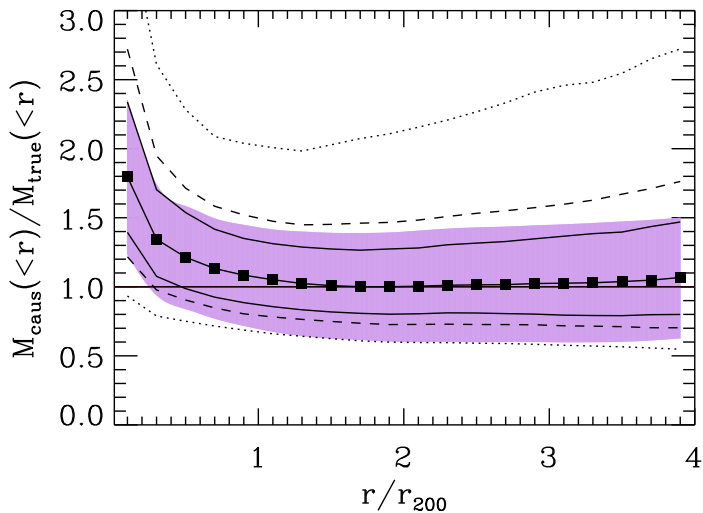}
\caption{Profiles of the ratio between the caustic and the true mass profile when the 
centre is forced to be the true one.
The lines and shaded areas are as in Figure \ref{fig:errors}.}
\label{fig:mass-ratio-rcenter}
\end{figure}

In Sect. \ref{sec:CGP}, we show how the cluster centre generally is well recovered.
The deviations are smaller than $0.07^\circ$ on the sky and 250~km~s$^{-1}$ along the line of sight in 90 percent
of the clusters. These deviations produce negligible effects on the mass
profile estimate. Figure \ref{fig:mass-ratio-rcenter} shows the ratio between
the caustic mass profile and the true mass profile, when the centre of the cluster is imposed to
be the true one. A comparison with Figure \ref{fig:mass-newF} 
shows that the average profile is still correct at radii larger than $0.6 r_{200}$: forcing the method 
to use the correct centre only slightly improves the estimate and reduces the scatter at 
radii larger than $\sim 2.5 r_{200}$.
It is worth pointing out that forcing the code to use the correct centre alters the quantities involved in the construction of the redshift diagram, specifically we recompute the velocity dispersion and the cluster size $R$ of the candidate cluster members with respect to the new centre. We then derive the new redshift diagram and locate the caustics. Despite these variations, the 
mass profile estimate does not substantially change. This result confirms the robustness of the method.

\subsection{Tuning the parameters}\label{sec:finetuning}

\begin{figure*}
\includegraphics[angle=0,scale=.57, bb=65 10 504 330]{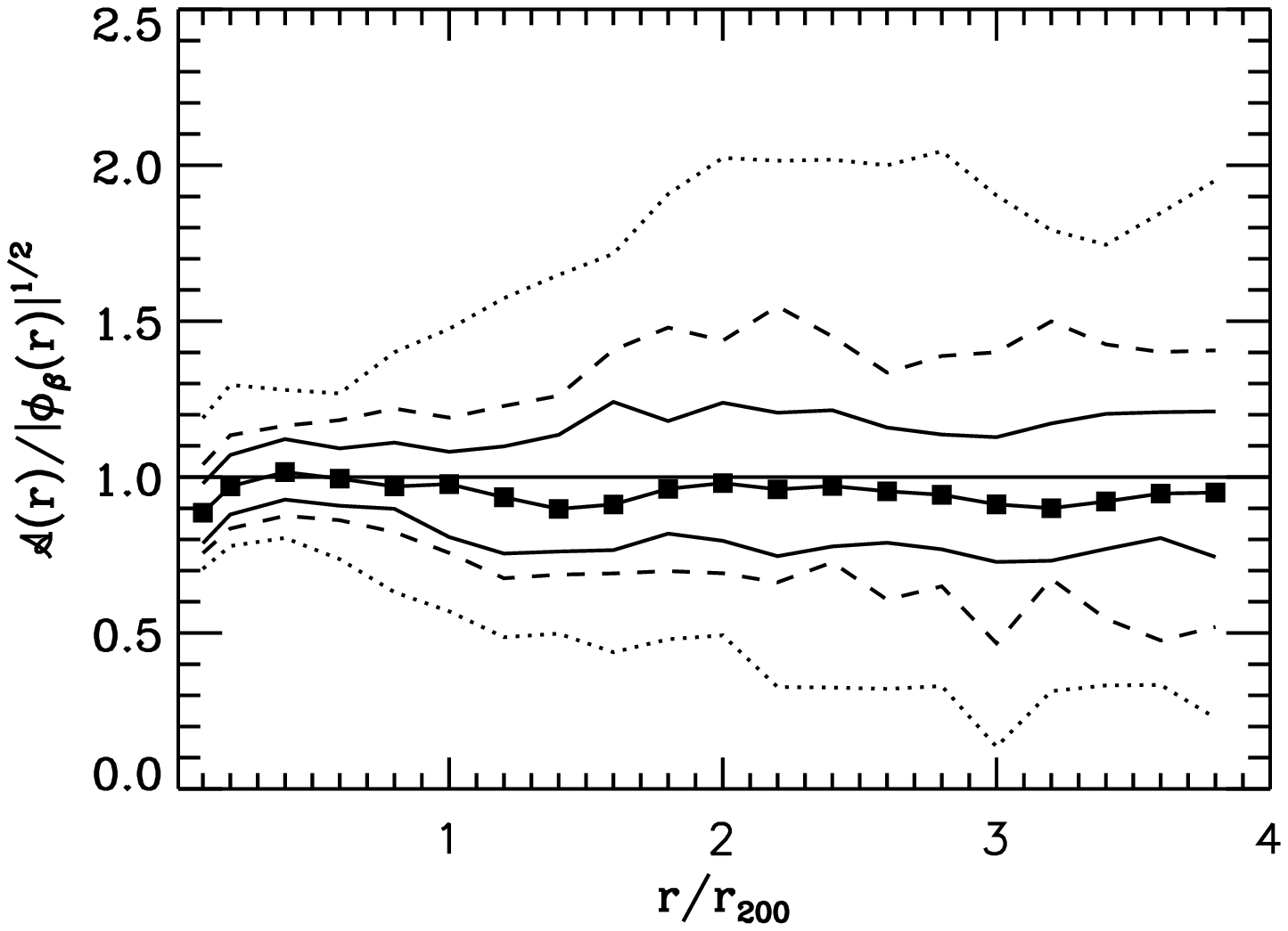}
\includegraphics[angle=0,scale=.57, bb=70 10 504 330]{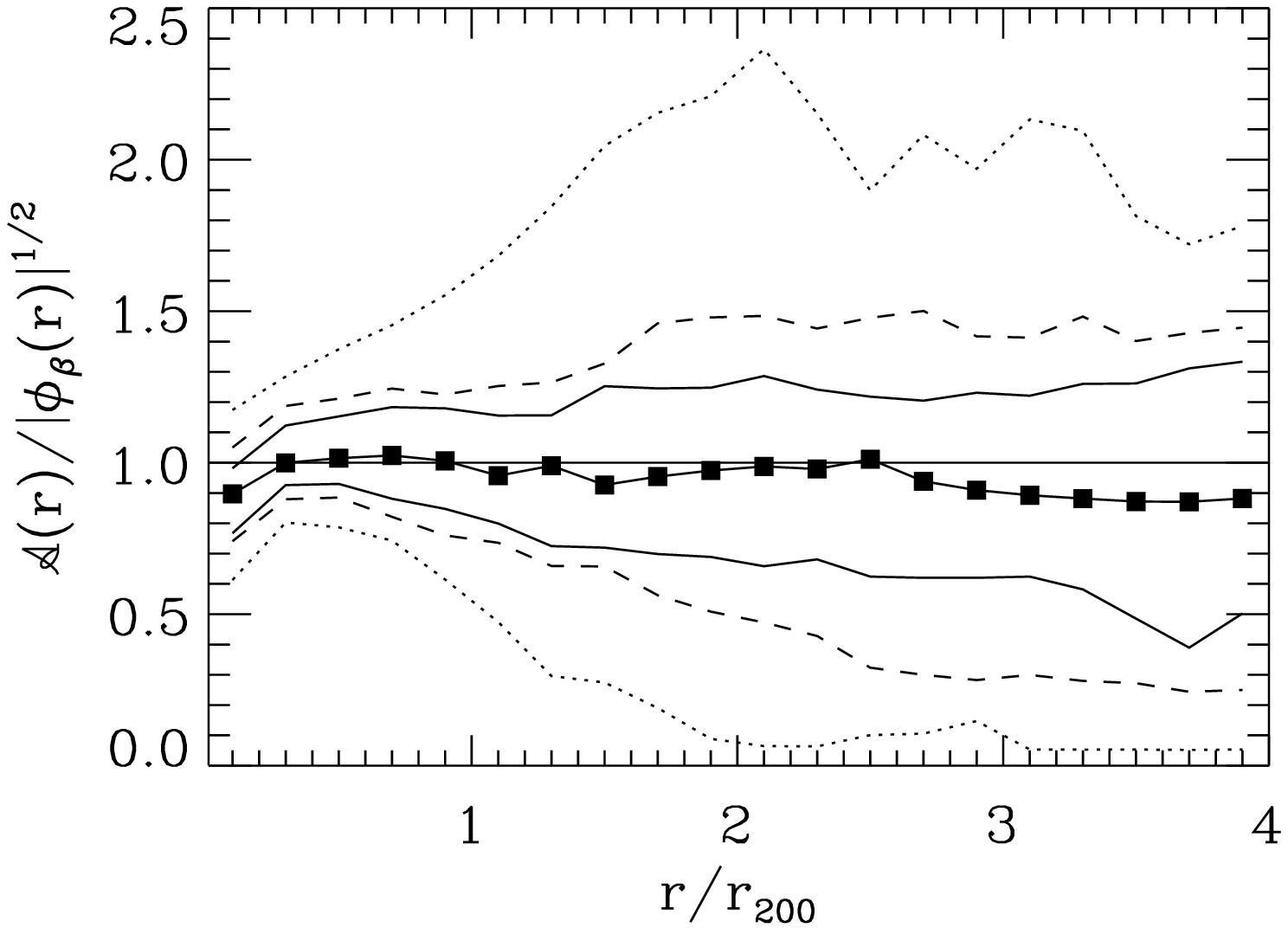}
\caption{Profiles of the ratio between the caustic and the true escape velocity 
profiles when the parameters are finely tuned (left panel) 
and from the automatic procedure (right panel) in a subsample of 100 mock catalogues: 
50, 68, and 90 percent of the profiles are within the upper and lower solid, dashed, and dotted curves,
respectively. The solid  squares show the median profiles.}
\label{fig:mass-5los}
\end{figure*}

The algorithm allows the user to choose some parameters, namely the 
thresholds of the binary tree, the optimal smoothing length $h_c$ for
the galaxy density distribution in the redshift diagram, and the threshold $\kappa$.
This freedom is necessary when the target cluster is in particularly crowded regions,
or the galaxy sampling is too sparse. In these cases, the algorithm is either
unable to locate the caustics or it returns unrealistic caustic amplitudes. 
If one is interested in the average profiles of a large cluster sample, tuning
the caustic parameters for each 
individual cluster can be very time consuming. To quantify the impact
of this possible fine tuning, for each of our 100 simulated clusters, we randomly draw one of the 30 
projections and tune the input parameters by hand until the caustic location appears
to be close enough to where we might expect them to be by eye. It turns out that most
clusters need a rather minor fine tuning, as demonstrated by the final result shown in Figure
\ref{fig:mass-5los}. The left-hand panel  shows the caustic amplitude when 
we apply a fine tuning to the parameters; the right-hand panel is for the
same sample without fine tuning. Clearly, the median profile remains unchanged,
but the scatter is slightly reduced.

\subsection{Projection effects}\label{sec:projeff}
\begin{figure*}
\includegraphics[angle=0,scale=.5, bb=70 10 504 330]{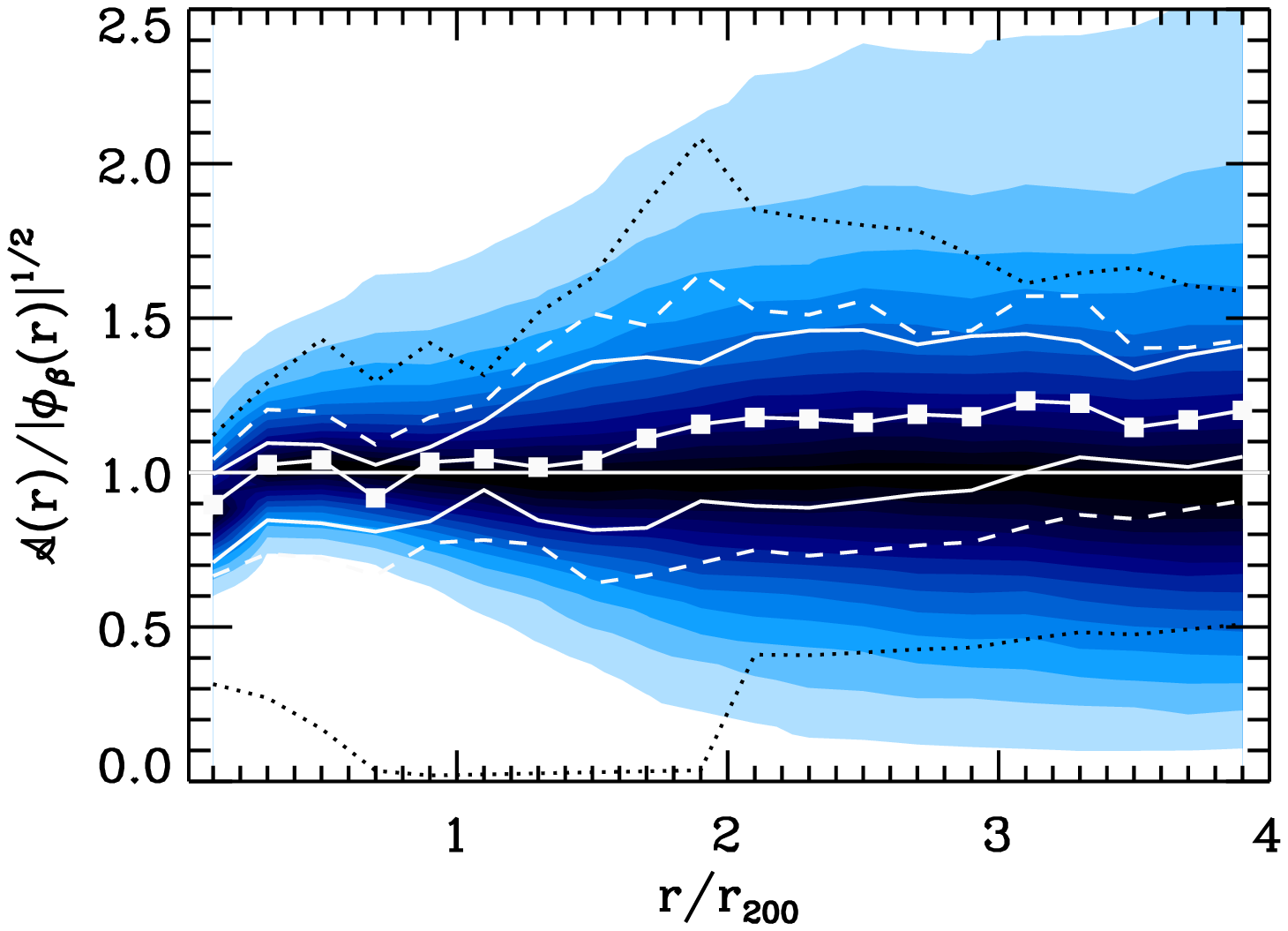}
\includegraphics[angle=0,scale=.5, bb=70 10 504 330]{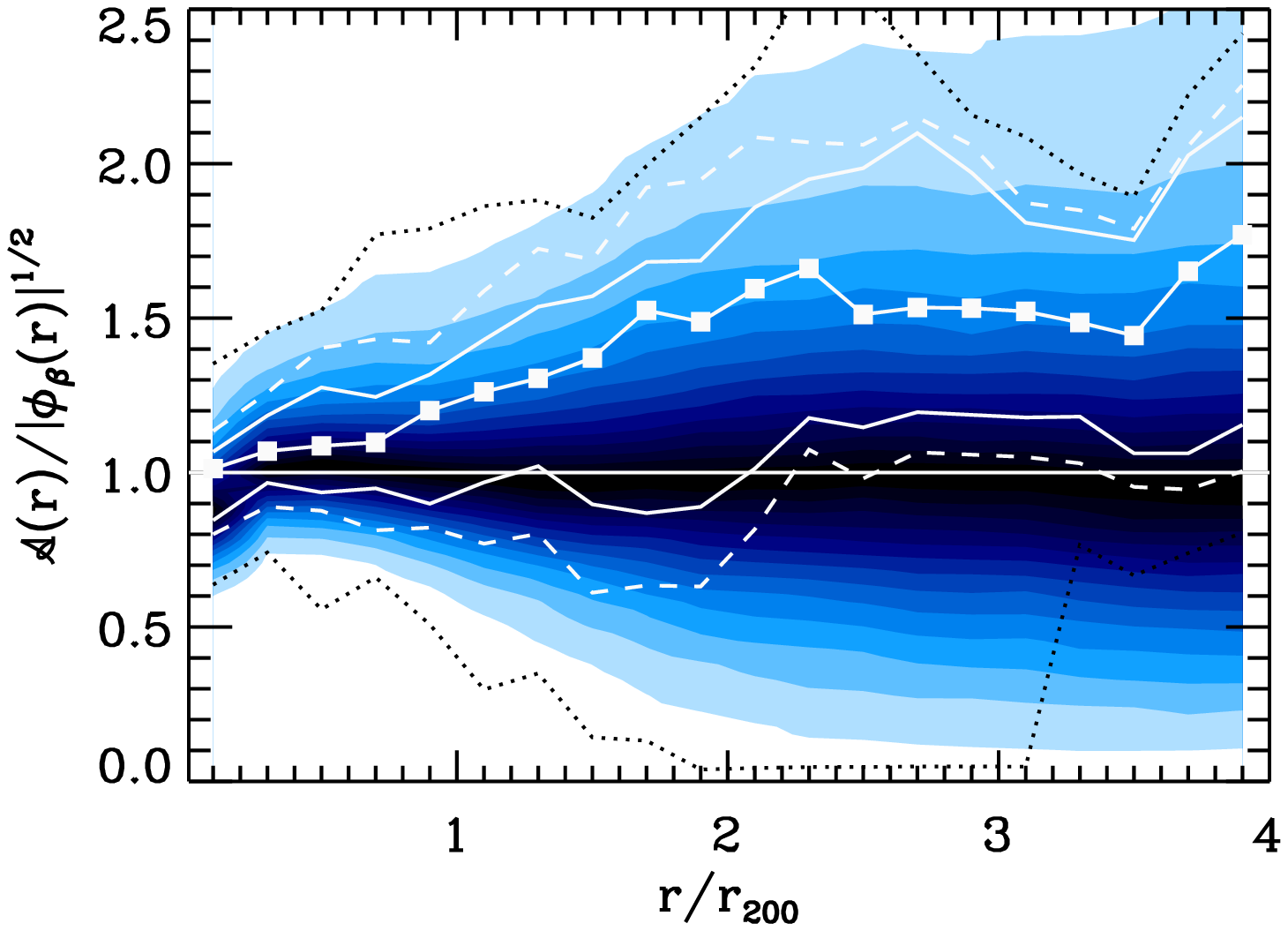}
\includegraphics[angle=0,scale=.5, bb=70 10 504 330]{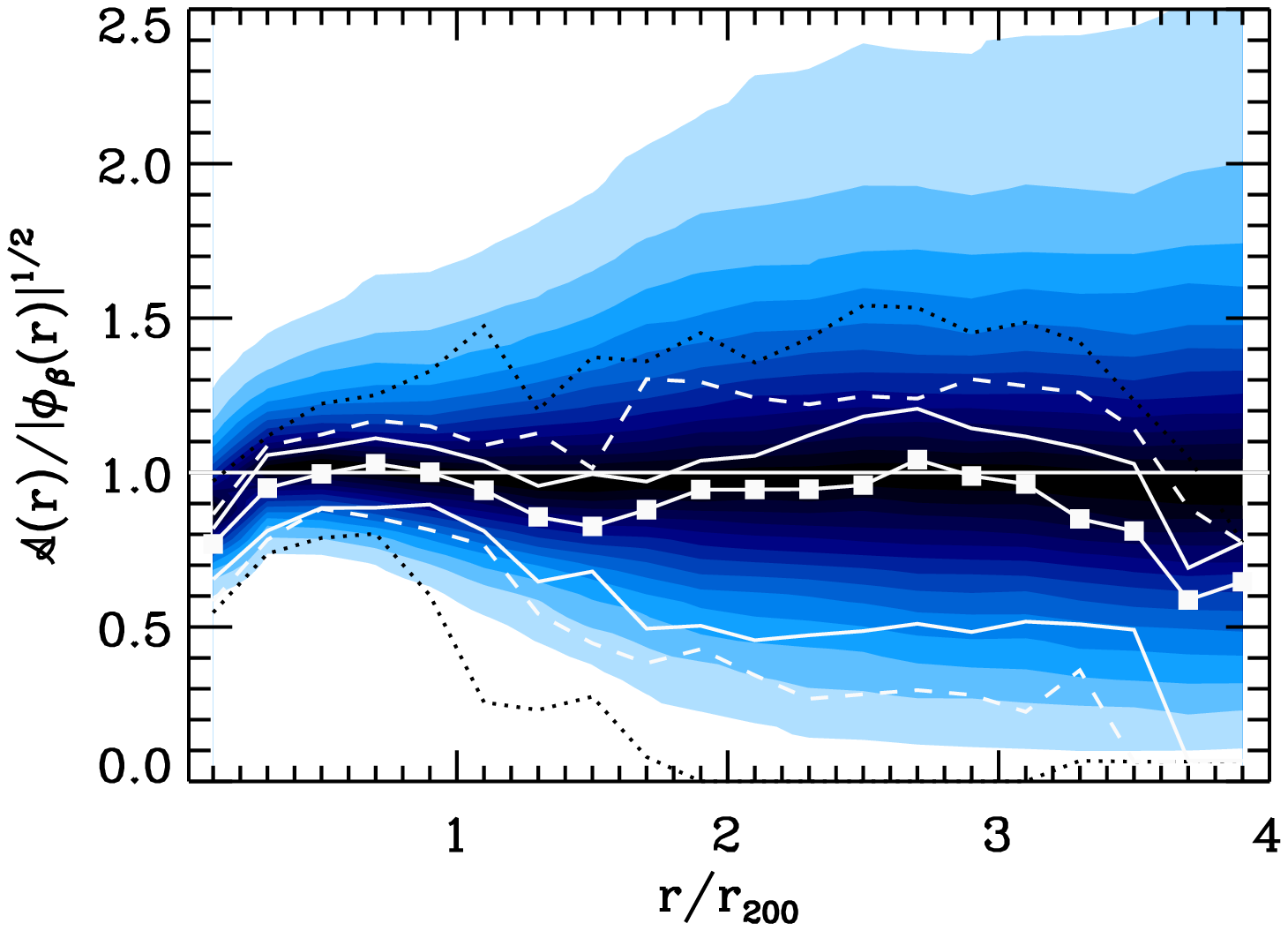}
\includegraphics[angle=0,scale=.5, bb=70 10 504 330]{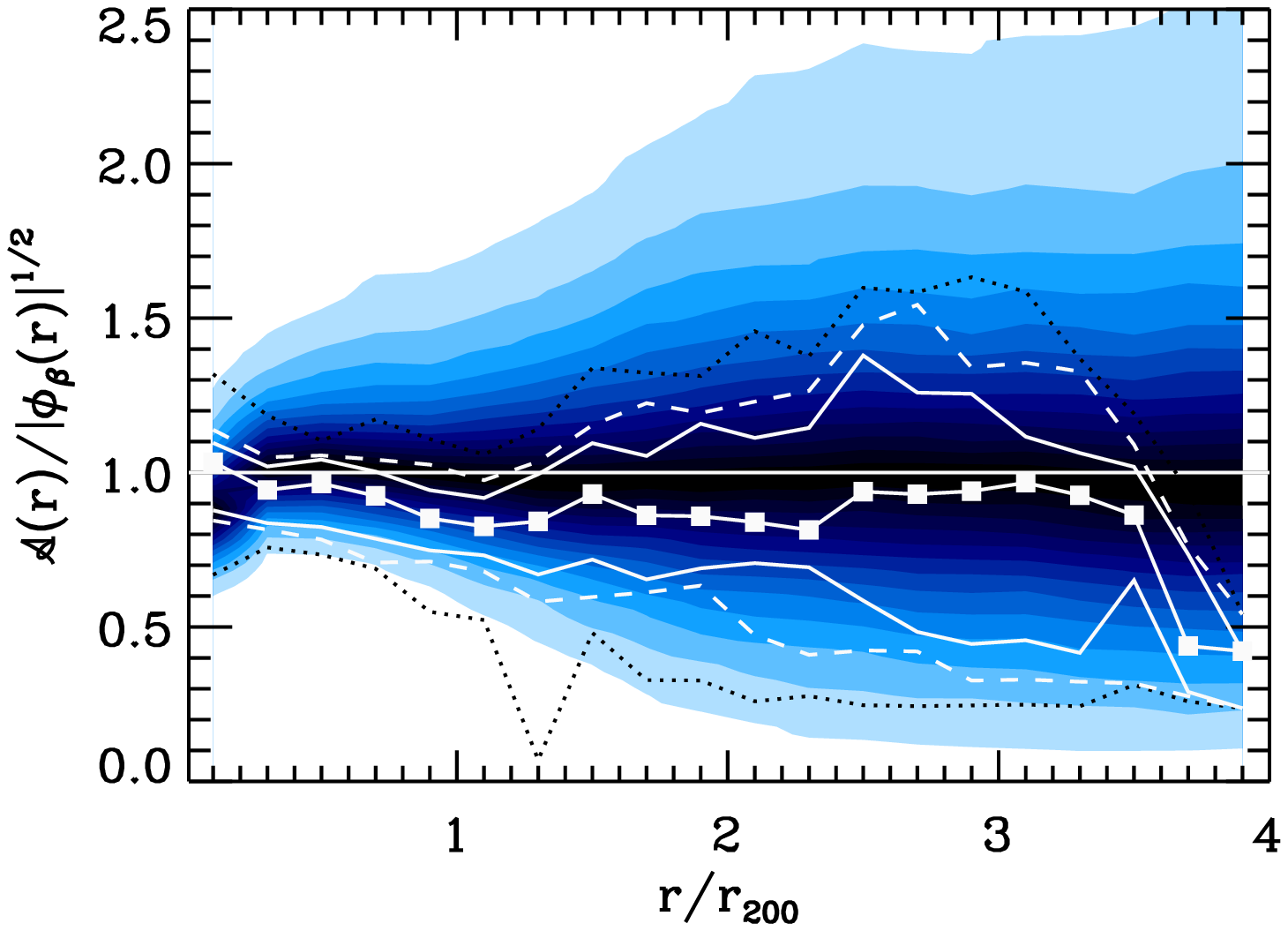}
\caption{Profiles of the ratio between the caustic amplitude ${\cal A}(r)$ and 
the l.o.s. component of the true escape velocity $\langle v_{\rm esc, los}^2(r)\rangle = -2\phi(r)/g(\beta)\equiv \phi_\beta(r)$ for four randomly chosen clusters.
The numerical gravitational potential profile $\phi(r)$ is
derived from the true mass distribution within $r_{\rm max}=10 h^{-1}$ Mpc from
the cluster centre: 50, 68, and 90 percent of the profiles derived from 30 different l.o.s. 
are within the upper and lower solid, dashed, and dotted curves,
respectively. The solid squares show the median profiles. The shaded areas are taken from Figure \ref{fig:caus-ratio}.  }
\label{fig:caus-ratio-ind}
\end{figure*}

The analysis provided above clearly shows that the uncertainties
on the cluster centre determination and the freedom on the
algorithm parameters are not responsible for most of the spread of
the escape velocity and mass profiles. 

This spread originates from the assumption of spherical symmetry.
In hierarchical clustering, this assumption does not hold in general,
but observationally our information is
limited to the galaxy distribution on the sky alone, 
although various techniques can in principle provide information on
the 3D shape of the cluster \citep[e.g.][]{zaroubi01, ameglio09}.

To show the impact of the projection effects on the caustic method, 
we take our 100 clusters and plot the escape velocity profiles derived from each of the 30 lines of sight. 
Figure \ref{fig:caus-ratio-ind} shows four randomly chosen clusters as examples.
The caustic technique 
returns a median profile systematically large for the 
cluster in the top-right panel. In the other three cases, however,
the median profile is within 30 percent of the correct one out to $\sim 3r_{200}$.

The relevant result of this test is the fact that the spread
due to the different lines of sight is comparable to the
spread of the entire sample (shaded area) taken from Figure \ref{fig:caus-ratio}.
This result clearly indicates that the projection effects 
are the major responsible for the systematic uncertainties
of the caustic technique and further refinements of the technique,
which still assume spherical symmetry,
appear to be unable to improve the mass estimate.

\subsection{Dependence of the profiles on the number of particles}\label{subsec:noofparticles}
\begin{figure*}
\centering
\includegraphics[angle=0,scale=.5,bb=30 20 470 360]{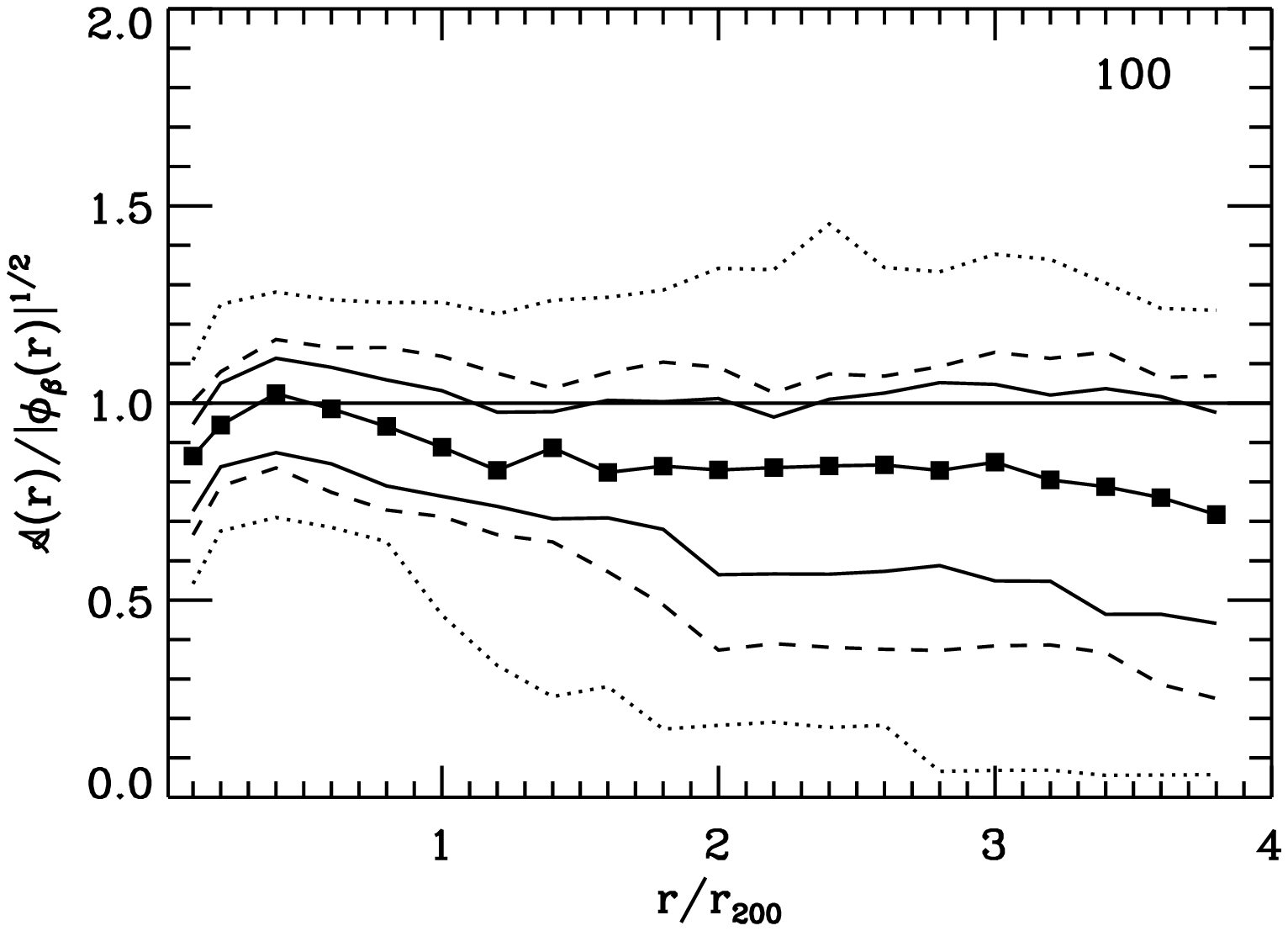}
\includegraphics[angle=0,scale=.5,bb=0 20 470 360]{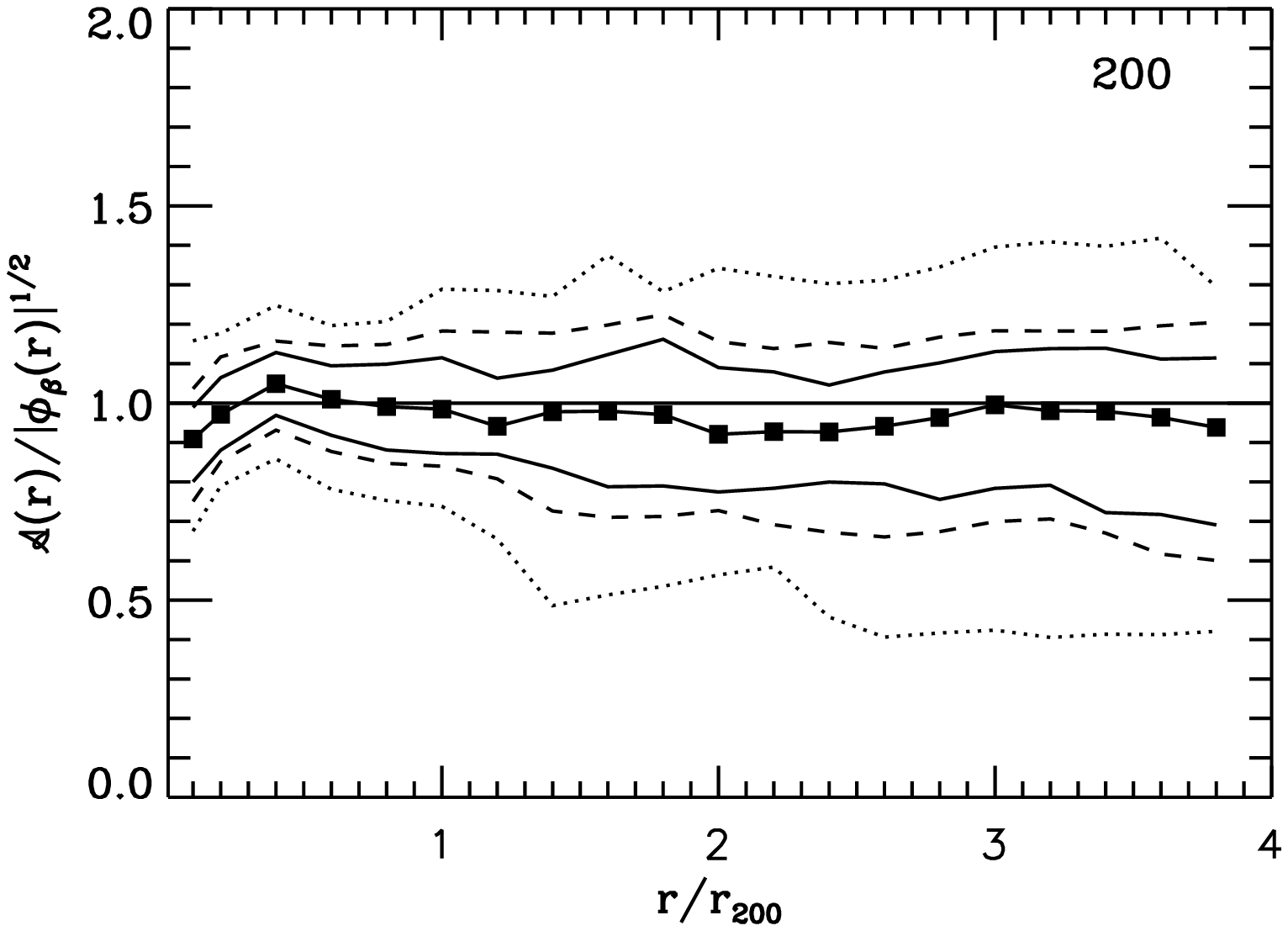}
\includegraphics[angle=0,scale=.5,bb=30 20 470 360]{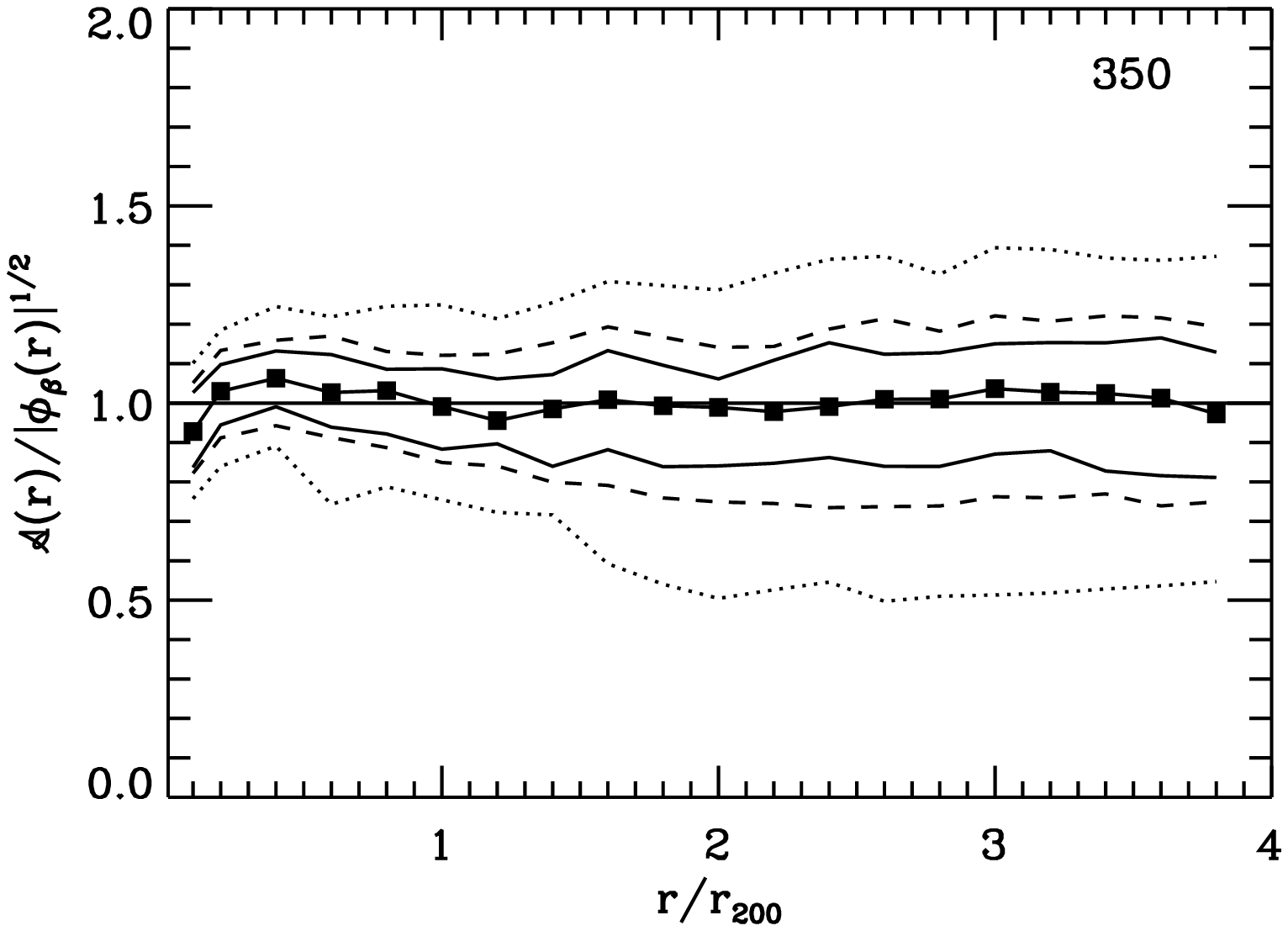}
\includegraphics[angle=0,scale=.5,bb=0 20 470 360]{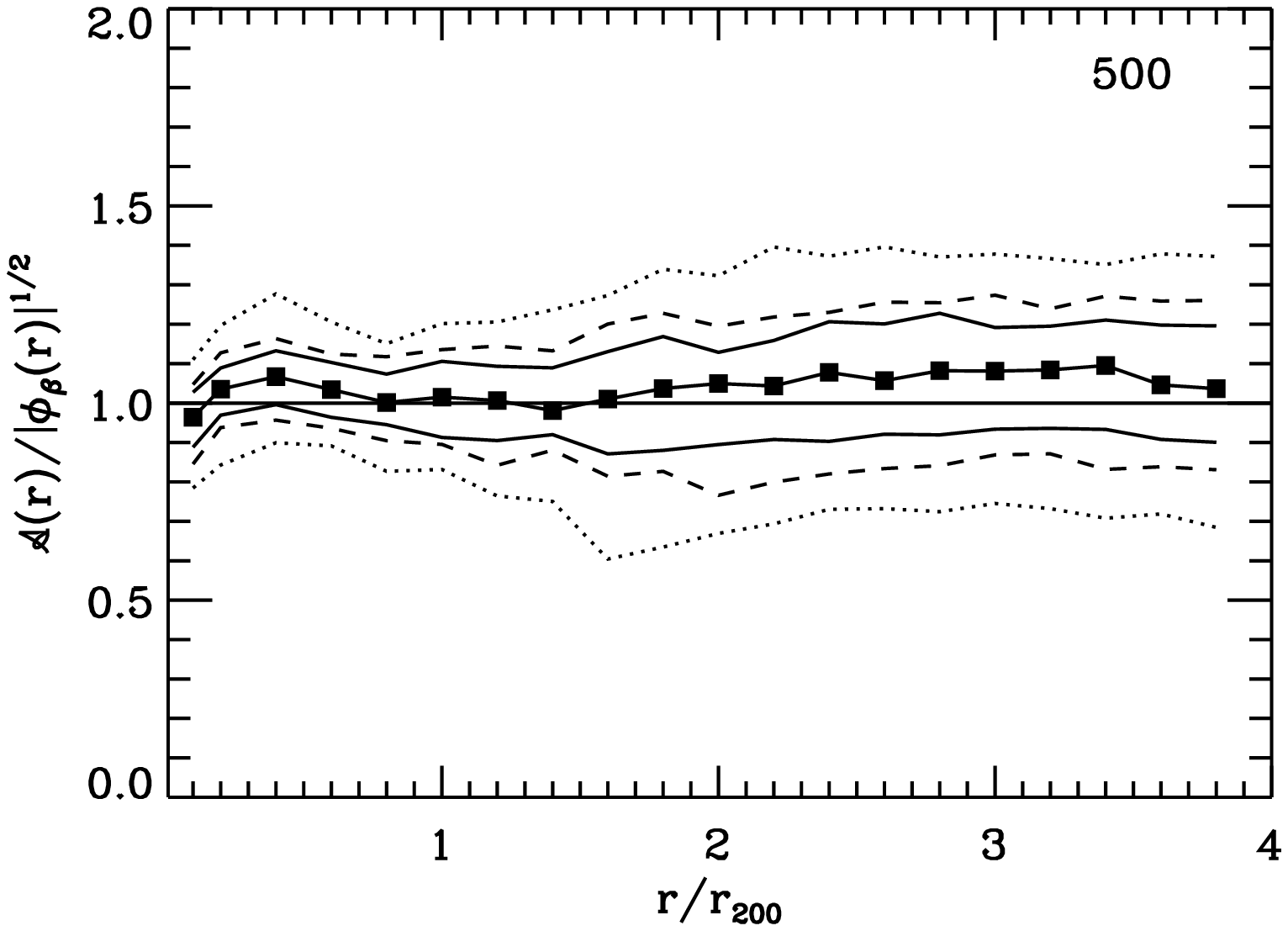}
\includegraphics[angle=0,scale=.5,bb=30 20 470 360]{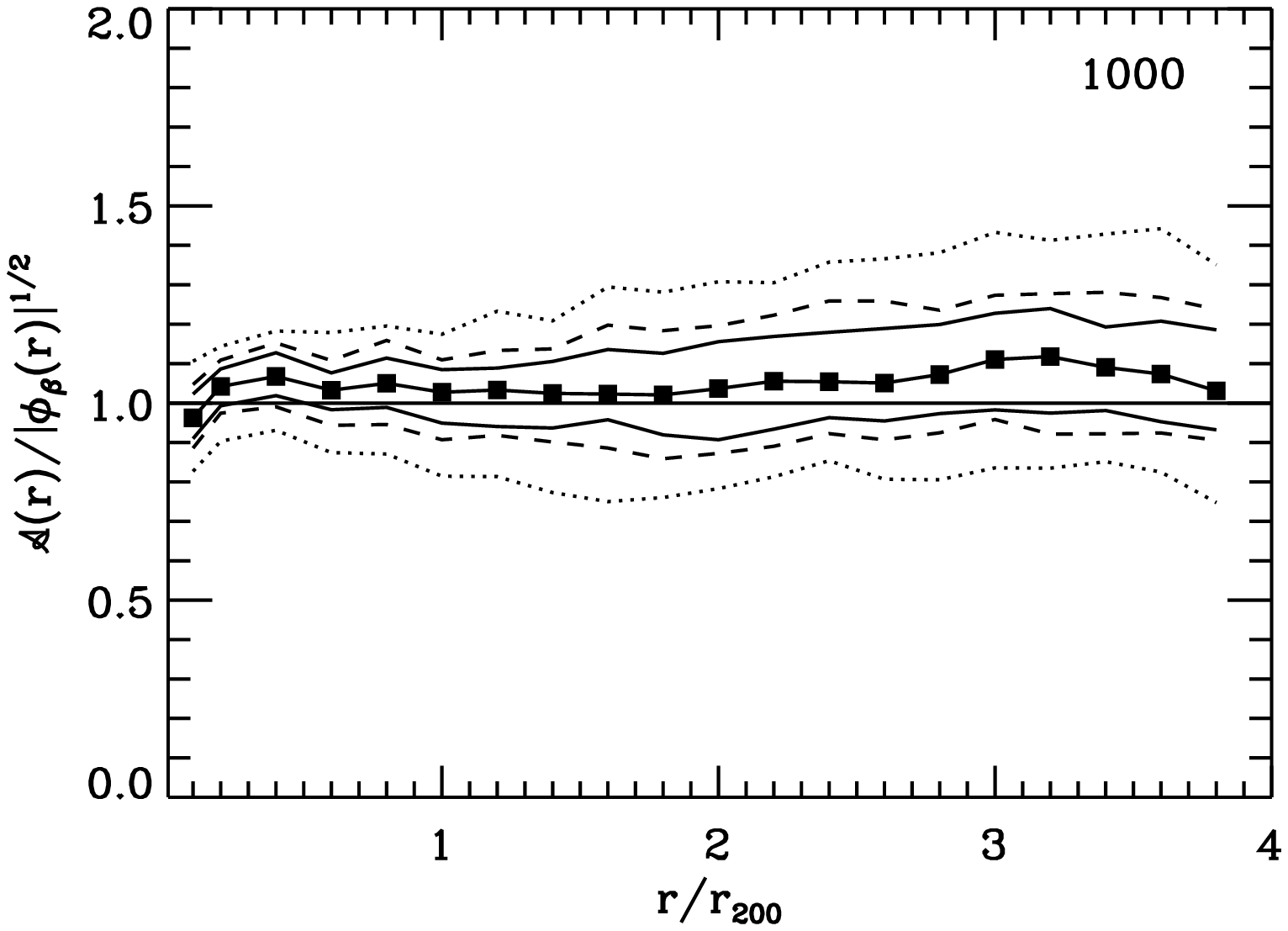}
\includegraphics[angle=0,scale=.5,bb=0 20 470 360]{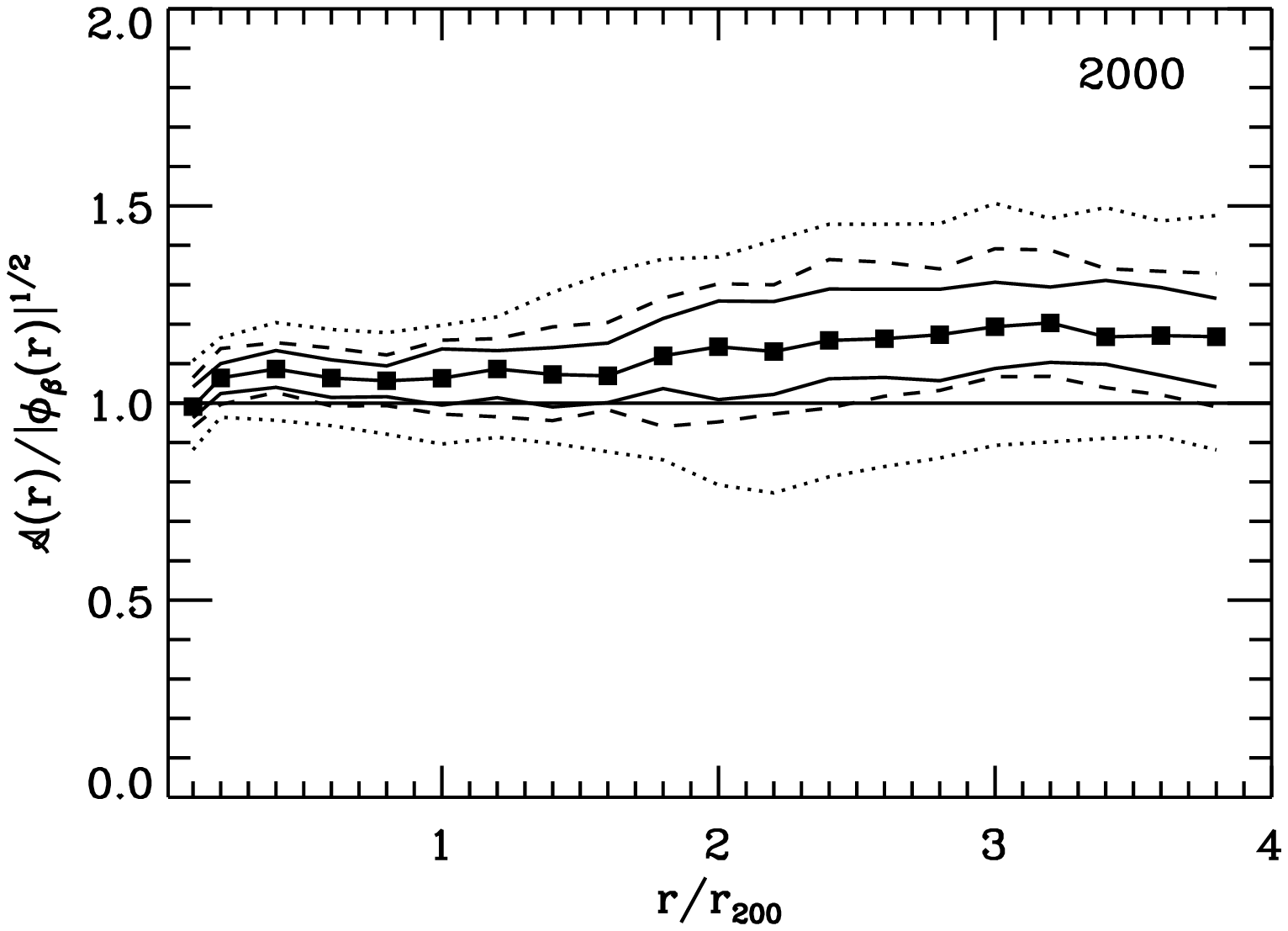}
\caption{Ratio between the estimated and true escape velocity profiles for subsamples with a given number
of particles within $3r_{200}$ as indicated in each panel;
50, 68, and 90 percent of the profiles are within the upper and lower solid, dashed, and dotted curves,
respectively. The solid squares show the median profiles.}
\label{fig:caus-ratio-2000}
\end{figure*}

To return a reliable estimate of the escape
velocity and mass profiles, the caustic technique requires a sufficient number of redshifts sampling 
the cluster velocity field.

To quantify this number, we stack our 3000 mock catalogues and apply the
caustic technique to the stacked cluster with an increasing number of
particles in the catalogue. With this procedure
we average over the different lines of sight and the final 
stacked cluster is very close to be spherically symmetric. 
We randomly choose particles in the 
catalogue until we obtain a given number $N$ of particles within 
$3r_{200}$ from the cluster centre in real space. We consider catalogues with
$N=[100, 200, 350, 500, 1000, 2000]$. We compile 100 different catalogues
for each value of $N$. The catalogues with the same $N$ clearly have
different total numbers of particles in the catalogue. Table \ref{table}  lists
the percentile ranges of the total number of particles in the catalogues,
which, as usual, have a field of view $12 \times  12$~$h^{-2}$~Mpc$^2$ at the cluster
redshift.
Table \ref{table} also lists the total number of particles in a field of view 
of $30^\prime\times 30^\prime$ ($2.46 \times  2.46$~$h^{-2}$~Mpc$^2$) and $|\Delta v_{\rm los}| \le 2000$~km~s$^{-1}$ 
centred on the cluster. 
 
Figure \ref{fig:caus-ratio-2000} shows the escape velocity profiles for the six different 
sets of catalogues. The spread decreases with the increasing number of particles,
as expected. However, the profile tends to be overestimated at large
radii and with richer catalogues.
This result originates from the fact that, as it was already
noticed by D99, \citet{schmalzing00}, and \citet{casa06}, mock redshift surveys
extracted from $N$-body simulations show large-scale structures
which are less sharp than in the real universe. Therefore, when 
we increase the number of particles in our mock catalogues,
we start sampling the surrounding structure of the
cluster in real space, the expected trumpet shape of the particle
density distribution in the redshift diagram becomes fuzzier, and the caustic algorithm has increasing difficulty at
locating the caustics properly. This problem is not present
in real surveys \citep[e.g.][]{rines06a}.
At these scales, the dark matter particles, that we use here, and the galaxies might not
share the same distribution in the redshift diagram
and an appropriate galaxy formation algorithm in mock surveys might provide
sharper large-scale structures. At any rate, Figure \ref{fig:caus-ratio-2000} shows 
that as few as $\sim 200$ galaxies within
$3r_{200}$, i.e. a few tens of redshifts per squared comoving megaparsec within the cluster,
are sufficient to have a reasonably accurate estimate of the escape velocity
profile out to $\sim 4r_{200}$.

\begin{table*}
\begin{tabular}{c|ccc|ccc}
\multicolumn{7}{|c|}{\bf Number of particles} \\
\hline
within $3r_{200}$&\multicolumn{3}{|c|}{in the catalogue}    &    \multicolumn{3}{|c|}{FOV} \\ 
$N$                 & 10\%    & median    & 90\%        & 10\%   & median    & 90\%        \\
\hline
100              &  406    &  470      &    556      & 89     &    98     &    106      \\
200              &  841    &  948      &   1023      & 184    &   194     &    209      \\
350              &  1504   &  1628     &   1772      & 326    &   339     &    355      \\
500              &  2162   &  2327     &   2477      & 465    &   487     &    506      \\
1000             &  4497   &  4693     &   4887      & 947    &   975     &   1003      \\
2000             &  9013   &  9359     &   9625      & 1904   &  1946     &   1979      \\

\end{tabular}

\caption{Number of particles in the catalogues compiled from the stacked cluster. 
The FOV columns list the numbers of particles within the field of
view $30^\prime\times 30^\prime$ ($2.46 \times 2.46$~$h^{-2}$~Mpc$^2$) and $|\Delta v_{\rm los}| \le 2000$~km~s$^{-1}$
centred on the cluster.}
\label{table}
\end{table*}

\section{Discussion and conclusion}\label{sec:concl}

We use a large sample of galaxy clusters extracted from a cosmological
$N$-body simulation to show that, by using galaxy redshifts alone, the
caustic technique measures (1) the escape velocity profiles of individual galaxy
clusters with 1-$\sigma$ uncertainty between $\sim 20$ and $\sim 50$ percent 
when the clustrocentric distances increase from $\sim 0.1 r_{200}$ to $4 r_{200}$, 
and (2) the cluster mass profiles with $\sim 50$ percent 1-$\sigma$ uncertainty
at clustrocentric distances in the range $\sim (0.6-4)r_{200}$.
The technique returns reliable estimates when a few tens of redshifts
per squared comoving megaparsec within the cluster are available. For
sparser samples, the technique can be applied to synthetic clusters
obtained by stacking individual clusters. The technique relies on the
kinematic properties of galaxy clusters in hierarchical clustering
scenarios of structure formation and assumes spherical symmetry. This
assumption alone is responsible for most of the estimated uncertainty.

The caustic amplitude measures the profile of the escape velocity
along the line of sight, which is a combination of the gravitational
potential profile and the velocity anisotropy parameter profile $\beta(r)$.  There are
extensive attempts to estimate $\beta(r)$ and mass profiles by
modelling the density of galaxies in the redshift diagrams with
sensible distribution functions \citep{wojtak09, wojtak10}. These
methods assume dynamical equilibrium and the caustic measure can thus
provide a robust consistency check of these estimates of $\beta(r)$
within $r_{200}$.

\citet{cupani08} have proposed a new approach, based on the spherical
collapse model, to estimate the mass profile of galaxy clusters beyond
their virial radius.  The spherical collapse model poorly describes
the formation of individual clusters in hierarchical clustering
scenarios. Nevertheless, \citet{cupani08}'s method seems to provide
cluster masses with $50$ percent uncertainty on average, once the mean
density of the universe $\Omega_0$ is set and an appropriate subsample
of galaxies is chosen based on their phase-space coordinates
\citep{cupani10}.

In addition to the caustic method and the \citet{cupani08}'s technique, the only method 
available to estimate the mass of clusters beyond their virial radius
is gravitational lensing.
However, the lensing signal is strong enough only when the cluster is within the redshift
range $z\sim 0.1-1$, whereas, in principle, the caustic technique can be applied to 
clusters at any redshift and it is only limited by the observing time 
required to measure a large enough number of galaxy spectra. 
The caustic technique relies on the assumption that
randomly chosen galaxies are fair tracers of the velocity field,
whereas gravitational lensing provides a direct measure of the projected
mass distribution.
All methods based on optical data relies on the absence
of velocity bias between galaxies and dark matter, and both $N$-body simulations
\citep[e.g.][]{diaf01, gill04, diemand04,
gill05} and observations \citep[e.g.][]{rines08}
indicate that this assumption is indeed reasonable.
Selected galaxy samples do however have velocity biases: in fact, elliptical galaxies, disk galaxies, 
and galaxies showing signs of interactions can have substantially 
different velocity distributions both in observations \citep[e.g.][]{biviano04, moss07}
and in simulated clusters \citep[e.g.][]{diaf99}.

The caustic technique estimates the escape velocity from a
cluster. Therefore, it can also be used to identify the galaxy members
of the cluster. This argument is not limited to clusters, and the
caustic technique was indeed applied to removing stellar interlopers
in dwarf galaxies \citep{ser09} and the Milky Way stellar halo
\citep{bro10}. The technique arranges the galaxies in the cluster
field in a binary tree and a by-product of this analysis is the
substructure identification.  Here, we have focussed on the estimate
of the escape velocity and mass profiles of galaxy clusters. We will
investigate the use of the caustic technique to identify members and
substructure of self-gravitating systems in future work.

With the advent of the future generation of large spectroscopic
surveys, both from ground (e.g., BigBOSS\footnote{\tt
  http://bigboss.lbl.gov/}) and from space (e.g., EUCLID\footnote{\tt
  http://sci.esa.int/euclid}), the application of the caustic technique
can represent an extremely valuable tool to estimate mass profiles
and galaxy membership over a large ensemble of galaxy clusters out to large
radii from the cluster centre, thus providing important
information on the cosmological assembly of dark matter halos
and on the galaxy-environment connection.

\section*{ACKNOWLEDGEMENTS}
AD warmly thanks Margaret Geller and Ken Rines whose contribution to the 
development and the diffusion of the caustic method has been invaluable 
over more than a decade now. AD also acknowledges Matthias Bartelmann and the Institut f\"ur
Theoretische Astrophysik (ITA) in Heidelberg for their warm
hospitality offered during the realization of part of this work.
We thank the referee for her/his careful reading of the manuscript
that helped to find some inaccuracies in our initial presentation of our results. We
acknowledge partial support from the INFN grant PD51,
the PRIN-MIUR-2008 grant ``Matter-antimatter asymmetry, dark matter 
and dark energy in the LHC Era'', the PRIN-MIUR-2007 grant ``The cycle of cosmic baryons'' and 
the contract ASI/COFIS I/016/07/0.

\bibliographystyle{mn2e}
%\bibliography{als}

 \end {document}